%% file: gbtw.tex
\documentclass[aps,prd,
noeprint,
preprint,
amsmath,
amssymb,
superscriptaddress,
nofootinbib
]{revtex4-1}

\pdfoutput=1
\usepackage[utf8]{inputenc}
\usepackage{graphicx}
\usepackage{subfig}
\usepackage{hyperref}
\usepackage{xcolor}
\usepackage{float}
\usepackage{pgffor}
\usepackage{multirow}
\usepackage{makecell}
\usepackage{booktabs}
\usepackage{enumitem}

\newcommand{\PreserveBackslash}[1]{\let\temp=\\#1\let\\=\temp}
\newcolumntype{C}[1]{>{\PreserveBackslash\centering}p{#1}}
\newcolumntype{R}[1]{>{\PreserveBackslash\raggedleft}p{#1}}
\newcolumntype{L}[1]{>{\PreserveBackslash\raggedright}p{#1}}
\newcommand\numberthis{\addtocounter{equation}{1}\tag{\theequation}}
\newcommand{\nn}{\nonumber \\}

\allowdisplaybreaks[2]

\newcommand{\SKLP}{State Key Laboratory of Particle Detection and Electronics, University of 
Science and Technology of China, Hefei 230026, Anhui, People’s Republic of China}
\newcommand{\USTC}{Department of Modern Physics, University of Science and Technology of China, 
Hefei 230026, Anhui, People’s Republic of China}

\begin{document}

\title{Two-loop master integrals for the single top production associated with $W$ boson}

\author{Ming-Ming Long}
\email{heplmm@mail.ustc.edu.cn}
\affiliation{\SKLP}
\affiliation{\USTC}

\author{Ren-You Zhang}
\email{zhangry@ustc.edu.cn}
\affiliation{\SKLP}
\affiliation{\USTC}

\author{Wen-Gan Ma}
\email{mawg@ustc.edu.cn}
\affiliation{\SKLP}
\affiliation{\USTC}

\author{Yi Jiang}
\email{jiangyi@ustc.edu.cn}
\affiliation{\SKLP}
\affiliation{\USTC}

\author{Liang Han}
\email{hanl@ustc.edu.cn}
\affiliation{\SKLP}
\affiliation{\USTC}

\author{Zhe Li}
\email{brucelee@mail.ustc.edu.cn}
\affiliation{\SKLP}
\affiliation{\USTC}

\author{Shuai-Shuai Wang}
\email{wang1996@mail.ustc.edu.cn}
\affiliation{\SKLP}
\affiliation{\USTC}

\date{\today}

\begin{abstract}
The $tW$ associated production has the second-largest cross section among three single top production channels. A complete study of NNLO QCD corrections to $tW$ production is still missing in literature. We present the calculation of part of the requisite two-loop master integrals for the process $b+g \to t+W$ at NNLO QCD. It turns out that the 80 master integrals in two families, involving up to three massive internal lines, can be expressed by Goncharov polylogarithms. The canonical differential equations that these integrals obey contain at most three square roots. By separating the canonical master integrals into groups according to their dependences on roots, we succeed in achieving optimization of final expressions, which are validated by independent numerical checks.
\begin{description}
\item[keywords]
Single Top Production, Master Integrals, Differential Equations
\end{description}
\end{abstract}

\maketitle

\input{./src/introduction}
\clearpage
\input{./src/topology}
\input{./src/deq.tex}

\section{\label{sec:summary}Summary}
In this paper, we present the computation of the two-loop MIs that contribute to the NNLO QCD corrections to the scattering process $b+g \to t+W$, the missing part of single top production at this order. We focus on one type of planar Feynman diagrams. From which two integral families, with up to three massive propagators, are identified. The total 80 MIs of these two families are calculated analytically by applying the differential equation method. We use the Magnus exponential method to build the canonical basis from the basis whose differential equation is linear in $\epsilon$. It turns out that at most, three square roots get involved in our problem. By noting that not every MI depends on all the roots, we find a way to optimize the final solutions, to compute MIs group by group according to their dependences on roots. In such a manner, the expressions of simple MIs would not be polluted by the change of variables that is needed to rationalize the roots for complicated MIs. Boundary conditions of the equation system are imposed either by requiring the finiteness of the solutions in some special kinematic limits (at pseudo-thresholds of MIs) or matching against the known integrals in the literature. Finally, the complete solutions in terms of GPLs up to weight 4 are provided. As a byproduct, we obtain the complete asymptotic expansions of all MIs in the limit $s\to0, t\to0,m_W\to0,m_t\to1$ when performing numerical checks of our analytic expressions.

To achieve the complete NNLO QCD corrections to single top production, we need to include more diagrams. However, other diagrams for $tW$ production possess more complex structures. It is expected, for example, that the elliptic sectors would come into play, which complicates the problem and deserves further investigations in the future.

\section{\label{sec:ac}Acknowledgements}
This work is supported in part by the National Natural Science Foundation of China (Grants No. 11775211 and No. 12061141005) and the CAS Center for Excellence in Particle 
Physics (CCEPP).

\bibliography{gbtw}

\end{document}

%% file: src/introduction.tex
\section{\label{sec:introduction}Introduction}

In the elementary particle spectrum of the Standard Model (SM), the top quark possesses the largest mass \cite{ParticleDataGroup:2020ssz}, comparable to that of a gold atom. That means its Yukawa coupling with the Higgs boson field is quite strong, indicating that top quark plays an essential role in the Electroweak Symmetry Breaking. Actually, the gluon fusion through a top loop dominates the Higgs production at the LHC. Another interesting feature of the top quark is that it decays before hadronization. Therefore, it is possible to measure top quark properties directly, considering that a huge amount of top quarks are produced at the LHC. Such a significant production rate ensures thorough investigations of the top quark, requiring higher-order theoretical predictions to test the SM at an unprecedented level.

Roughly speaking, the top quark production at the LHC can be classified into two categories, the pair and single production. One can further identify three channels in the single top production mode, namely the $t$-channel, the $tW$ associated production, and the $s$-channel, in the order of descending cross section at the LHC. Unlike the pair production, which is dominated by pure strong interaction, weak interaction comes into play in the single production mode, thereby providing an opportunity to probe the $W-t-b$ couplings \cite{ATLAS:2017ygi,ATLAS:2017rcx} and directly measure the Cabibbo-Kobayashi-Maskawa (CKM) matrix element $V_{tb}$ \cite{ATLAS:2019hhu,CMS:2020vac}. Besides its advantages in studying the electroweak properties of the top quark, the single top production can also be used to measure top quark mass \cite{CMS:2017mpr,CMS:2021jnp}.

Plentiful studies available in the literature help construct a reliable understanding about the single top production. For the $t$- and $s$-channel, the next-to-leading order (NLO) QCD corrections at different levels have been available for years \cite{Bordes:1994ki,Stelzer:1997ns,Cao:2005pq,Campbell:2009ss,Smith:1996ij,Cao:2004ap,Heim:2009ku,Harris:2002md,Sullivan:2004ie,Campbell:2004ch,Cao:2004ky} and are finally implemented into fully flexible Monte Carlo programs \cite{Frixione:2005vw,Alioli:2009je}. Remarkably, the next-next-to-leading order (NNLO) QCD corrections to $t$-channel single top production have been obtained in Ref. \cite{Brucherseifer:2014ama} where the structure-function method is applied. Progress is made to include the decay of top quark under narrow width approximation in Refs. \cite{Berger:2016oht,Berger:2017zof,Campbell:2020fhf} and the calculation for $s$-channel is given in Ref. \cite{Liu:2018gxa} as well. The structure-function approximation applied in the previous studies bypasses the computation of complicated two-loop four-point Feynman integrals involving too many scales. The complete analytic expressions of those integrals are still missing, while their numerical evaluations through the auxiliary mass flow method are provided in Ref. \cite{Bronnum-Hansen:2021pqc}.

On the other hand, as a complementary manner to investigate top quarks, the $tW$ channel deserves a detailed examination. The calculations at NLO QCD are presented in Refs. \cite{Tait:1999cf,Zhu:2002uj,Campbell:2005bb,Cao:2008af} and, again, included in Monte Carlo implementations \cite{Frixione:2008yi,Re:2010bp}. At higher orders, only approximate corrections based on an expansion of the results from soft gluon resummation calculations are provided in Refs. \cite{Kidonakis:2006bu,Kidonakis:2007ej} and Refs. \cite{Kidonakis:2010ux,Kidonakis:2015wva,Kidonakis:2016sjf,Kidonakis:2021vob} at the next-to-leading logarithm and next-next-to-leading logarithm level, respectively. To complete the NNLO QCD corrections to single top production, it is indisputable to include the exact missing higher-order effects for the $tW$ channel.

One of the technical issues in a complete NNLO QCD corrections to $tW$ production is the computation of the complex two-loop Feynman integrals. The relevant Feynman diagrams turn out to involve multiple scales and contain quite distinct structures, increasing the difficulty of handling the problem. It is natural to compute the related integrals group by group. The authors in Ref. \cite{Chen:2021gjv} present partial results where they separate the diagrams according to the number of massive propagators, and show that the master integrals (MIs) corresponding to planar and non-planar double-box diagrams with one massive propagator could be expressed in terms of Goncharov polylogarithms. In this paper, we will illustrate that some more complicated integrals involving up to three massive propagators are able to evaluate to Goncharov polylogarithms as well.

The remaining part of the paper is organized as follows. First, the classification of the planar Feynman diagrams that contribute to $tW$ production is discussed in section \ref{sec:diagrams}. We focus on one of the three types of planar diagrams and present the notations and conventions for the two integral families, which are identified from the diagrams we are considering, in the same section. Section \ref{sec:deq} constitutes the main content of this paper. In there we present the construction of canonical equations and show how the solutions to the equations can be expressed in terms of Goncharov polylogarithms once the relevant square roots are rationalized. Optimization of the solutions is realized by computing the MIs group by group according to their dependences on roots. Finally, a brief summary is given in section \ref{sec:summary}.

%% file: src/topology.tex
\section{\label{sec:diagrams}Two-loop four-point Feynman diagrams}

\begin{figure}[h]
  \centering
  \captionsetup[subfigure]{labelformat=empty}
  \subfloat[$\mathcal{T}_1$]{%
    \includegraphics[width=0.3\textwidth]{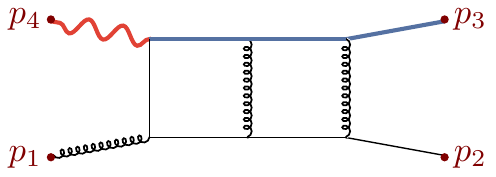}
  }
  \subfloat[$\mathcal{T}_2$]{%
    \includegraphics[width=0.3\textwidth]{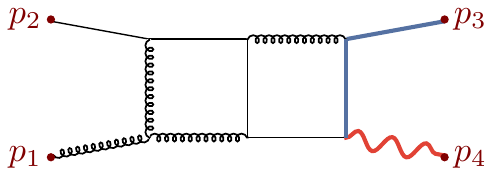}
  }
  \subfloat[$\mathcal{T}_3$]{%
    \includegraphics[width=0.3\textwidth]{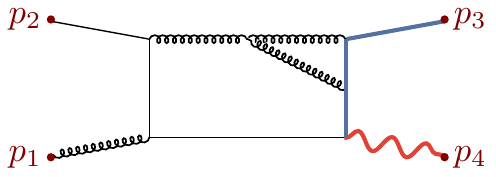}
  }\\
  \subfloat[$\mathcal{T}_4$]{%
    \includegraphics[width=0.3\textwidth]{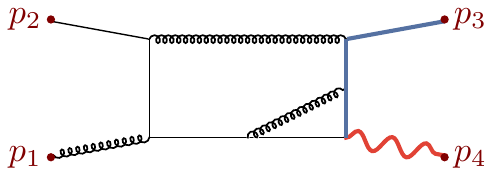}
  }
  \subfloat[$\mathcal{T}_5$]{%
    \includegraphics[width=0.3\textwidth]{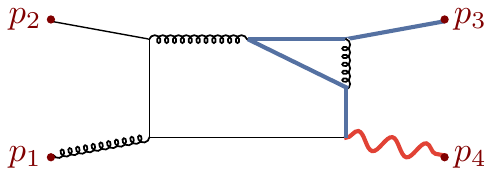}
  }
  \subfloat[$\mathcal{T}_6$]{%
    \includegraphics[width=0.3\textwidth]{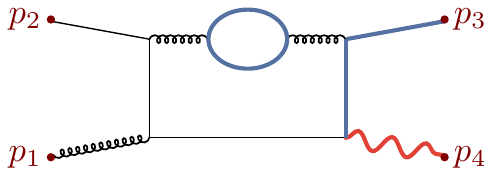}
  }
\caption{Sample diagrams of 1-2-3-4 type. The momenta are in the sequence $p_1$-$p_2$-$p_3$-$p_4$ clockwise or counterclockwise from the lower-left corner.}
 \label{fig:dia1}
\end{figure}

\begin{figure}[h]
  \centering
  \captionsetup[subfigure]{labelformat=empty}
  \subfloat[$\mathcal{T}_7$]{%
    \includegraphics[width=0.3\textwidth]{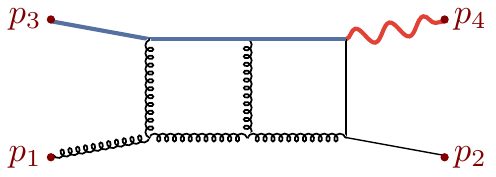}
  }
  \subfloat[$\mathcal{T}_8$]{%
    \includegraphics[width=0.3\textwidth]{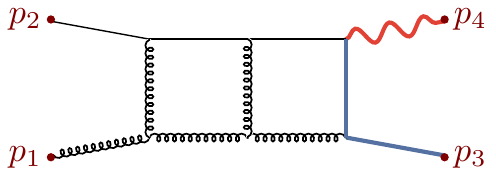}
  }
  \subfloat[$\mathcal{T}_9$]{%
    \includegraphics[width=0.3\textwidth]{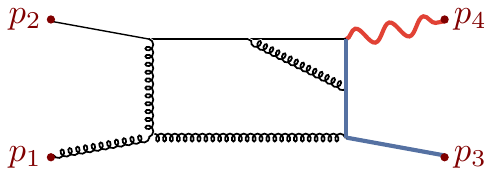}
  }\\
  \subfloat[$\mathcal{T}_{10}$]{%
    \includegraphics[width=0.3\textwidth]{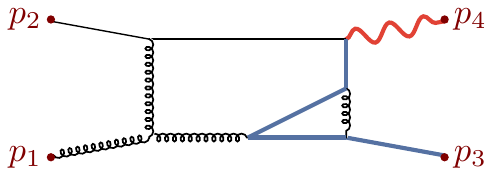}
  }
  \subfloat[$\mathcal{T}_{11}$]{%
    \includegraphics[width=0.3\textwidth]{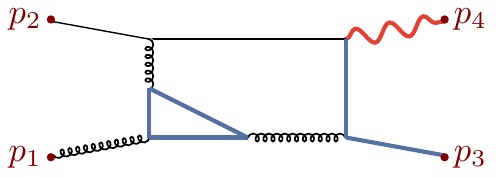}
  }
  \subfloat[$\mathcal{T}_{12}$]{%
    \includegraphics[width=0.3\textwidth]{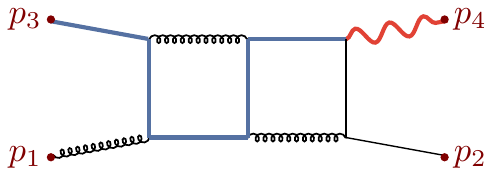}
  }
\caption{Sample diagrams of 1-2-4-3 type. The momenta are in the sequence $p_1$-$p_2$-$p_4$-$p_3$ clockwise or counterclockwise from the lower-left corner.}
 \label{fig:dia2}
\end{figure}

\begin{figure}[h]
  \centering
  \captionsetup[subfigure]{labelformat=empty}
  \subfloat[$\mathcal{T}_{13}$]{%
    \includegraphics[width=0.3\textwidth]{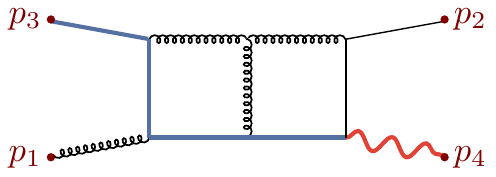}
  }
  \subfloat[$\mathcal{T}_{14}$]{%
    \includegraphics[width=0.3\textwidth]{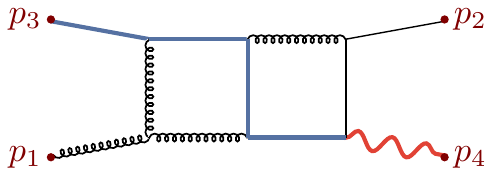}
  }
  \subfloat[$\mathcal{T}_{15}$]{%
    \includegraphics[width=0.3\textwidth]{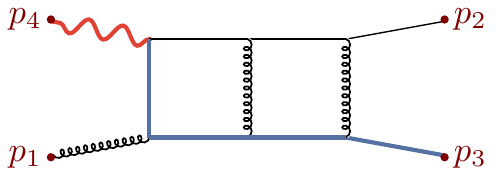}
  }\\
  \subfloat[$\mathcal{T}_{16}$]{%
    \includegraphics[width=0.3\textwidth]{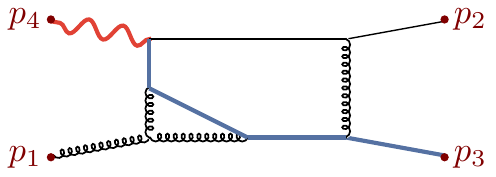}
  }
  \subfloat[$\mathcal{T}_{17}$]{%
    \includegraphics[width=0.3\textwidth]{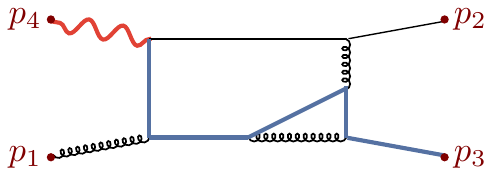}
  }
  \subfloat[$\mathcal{T}_{18}$]{%
    \includegraphics[width=0.3\textwidth]{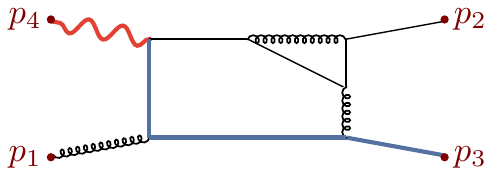}
  }
\caption{Sample diagrams of 1-3-2-4 type. The momenta are in the sequence $p_1$-$p_3$-$p_2$-$p_4$ clockwise or counterclockwise from the lower-left corner.}
 \label{fig:dia3}
\end{figure}

We consider the Feynman integrals that contribute to the NNLO QCD corrections to the partonic scattering process 
\begin{equation}
	g(p_1) + b(p_2) \to t(p_3) + W(p_4),
\end{equation}
with $p_{1,2}^2=0$, $p_3^2=m_t^2$ and $p_4^2=m_W^2$. The bottom quark is regarded as a massless particle in our calculation. Here, $m_t$ and $m_W$ stand for the masses of top quark and $W$ boson, respectively. And we define three Mandelstam invariants as
\begin{equation}
	s=(p_1+p_2)^2, ~~~
	t=(p_1-p_4)^2, ~~~
	u=(p_2-p_4)^2,
\end{equation}
with $s+t+u=m_t^2+m_W^2$. In many cases, the Feynman integrals from diagrams related to each other by swapping two of the four momenta share the same expressions up to different input parameters. However, that is not always true for our problem. The planar two-loop four-point Feynman diagrams for this process can be classified into three types according to the arrangement of the momenta $p_{1,2,3,4}$. They are the 1-2-3-4 type (Figure \ref{fig:dia1}), 1-2-4-3 type (Figure \ref{fig:dia2}) and 1-3-2-4 type (Figure \ref{fig:dia3}). It is clear that, for instance, the integrals from diagrams $\mathcal{T}_{11,12}$ in Figure \ref{fig:dia2} have new structures compared with Figure \ref{fig:dia1}. We limit ourselves to the calculation of Feynman integrals for diagrams in Figure \ref{fig:dia1} and leave others to the future.

\subsection{\label{sec:topology}Integral families}
We are going to compute the MIs related to the planar two-loop four-point Feynman diagrams, shown in Figure \ref{fig:dia1}, which contribute to the NNLO QCD corrections to $tW$ production at the hadron colliders. From these diagrams, two integral families can be identified:
\begin{itemize}
	\item The first family includes the diagrams $\mathcal{T}_{1,2,3,4}$ in Figure \ref{fig:dia1} with two massive propagators. It is defined by
\begin{align}\begin{alignedat}{3}
D_1 &= l_1^2, &                D_2 &= (l_1+p_1)^2, & D_3 &= (l_1+p_1+p_2)^2, \\
D_4 &= (l_2+p_1+p_2)^2,\quad & D_5 &= (l_2+p_4)^2-m_t^2,\quad& D_6 &= l_2^2, \\
D_7 &= (l_1-l_2)^2,   &     D_8 &= (l_1+p_4)^2-m_t^2,  &  D_9 &= (l_2+p_1)^2.
\end{alignedat}\end{align}
	\item The second family contains the diagrams $\mathcal{T}_{5,6}$ in Figure \ref{fig:dia1} with three massive propagators. It is defined by
\begin{align}
\begin{alignedat}{3}
D_1 &= l_1^2, &                D_2 &= (l_1+p_1)^2, & D_3 &= (l_1+p_1+p_2)^2, \\
D_4 &= (l_2+p_1+p_2)^2-m_t^2,\quad & D_5 &= (l_2+p_4)^2,\quad& D_6 &= (l_1+p_4)^2-m_t^2, \\
D_7 &= (l_1-l_2)^2-m_t^2,   &  D_8 &= l_2^2,     &  D_9 &= (l_2+p_1)^2,
\end{alignedat}
\end{align}
\end{itemize}
where we use $l_{1,2}$ to represent loop momenta. The two-loop MIs to be evaluated later in $d=4-2\epsilon$ dimensions have the form
\begin{equation}
	F(\alpha_1,...,\alpha_9) = \int \mathcal{D}^d l_1 \mathcal{D}^d l_2 \frac{1}{D_1^{\alpha_1}...D_9^{\alpha_9}}, ~~ \alpha_i \in \mathbb{Z},
\end{equation}
where $D_{1,...,9}$ could be propagators of family 1 or 2. And the integration measure is defined as
\begin{equation}
	\mathcal{D}^d l_i=\frac{(m_t^2)^{\epsilon}d^dl_i}{(2\pi)^d}\left( \frac{i S_{\epsilon}}{16\pi^2} \right)^{-1},
\label{eq:measure}
\end{equation}
with
\begin{equation}
	S_{\epsilon}=(4\pi)^{\epsilon}\Gamma(1+\epsilon).
\end{equation}
In such a convention, one of the MIs is normalized to be 1. 

The Feynman integrals in one family are not independent of each other. Linear relations between them indicate that a basic set of integrals, i.e. MIs, could be identified once the selecting regulations are fixed. Such relations could be provided by the integration-by-parts (IBP) identities \cite{Chetyrkin:1981qh}. An algorithm to find out the MIs based on Ref. \cite{Laporta:2001dd} has been encoded into efficient programs. We use \texttt{Kira} \cite{Maierhoefer:2017hyi,Klappert:2020nbg} to conduct the IBP reduction and find that the number of MIs is 60 for family 1 and 36 for family 2. Then what remains is the computation of the found MIs, which is the primary purpose of this paper.

%% file: src/deq.tex
\section{\label{sec:deq}Differential equations and solutions}

The differential equation method \cite{Kotikov:1990kg,Remiddi:1997ny} turns out to be a powerful tool to compute multi-loop Feynman integrals. Especially for the ones with many scales involved, like our case. Because they are too complicated to be calculated in a direct integration manner. While through differential equation method, very compact analytic expressions of multi-loop Feynman integrals could be obtained if a canonical basis exists in particular \cite{Henn:2013pwa}.

Let us write $\vec{\kappa}=(s,t,m_W^2,m_t^2)$ as the kinematic variable vector and denote the vector of MIs in a family by $\vec{\text{F}}(\vec{\kappa},\epsilon)$. One can establish a set of first-order differential equations w.r.t. each kinematic variable,
\begin{equation}
	\partial_i\, \vec{\text{F}}(\vec{\kappa},\epsilon) = B_i(\vec{\kappa};\epsilon)\,\vec{\text{F}}(\vec{\kappa},\epsilon),
\end{equation}
where $\partial_i=\partial/\partial \kappa_i$. Usually, the coefficient matrices $B_i(\vec{\kappa};\epsilon)$ have complex coupled dependences on kinematic variables and regulator $\epsilon$, making it difficult to solve the equation system. Suppose a new vector of MIs $\vec{\text{g}}$, which is related to $\vec{\text{F}}$ by a linear transformation $\vec{\text{g}}=\text{T}\,\vec{\text{F}}$, then the new equation system reads
\begin{equation}
	\partial_i\, \vec{\text{g}}(\vec{\kappa},\epsilon) = \epsilon\, A_i(\vec{\kappa})\,\vec{\text{g}}(\vec{\kappa},\epsilon),
\label{eq:epsform}
\end{equation}
where the new coefficient matrices are related to original ones by
\begin{equation}
	\epsilon\, A_i = \text{T}\,B_i\,\text{T}^{-1} - \text{T}\,(\partial_i\,\text{T}^{-1}).
\end{equation}
We can see that the regulator $\epsilon$ in Eq. (\ref{eq:epsform}) is factorized out. Furthermore, the matrices $A_i$ can be integrated into one matrix $\mathbb{A}$, whose entries are only the linear combinations of logarithms with rational number coefficients. Then the differential equation system Eq. (\ref{eq:epsform}) can be cast into the so-called $d\log$ form,
\begin{equation}
	d\, \vec{\text{g}}(\vec{\kappa},\epsilon) = \epsilon\,\left(\sum_i \mathbb{C}_i d\log(\beta_i(\vec{\kappa}))\right)\,\vec{\text{g}}(\vec{\kappa},\epsilon),
\label{eq:dlogform}
\end{equation}
where the system's kinematic dependences are encoded into the alphabet $\Omega=\{\beta_i\}$, and every $\beta_i$ is called a letter which is, generally, an algebraic function of $\vec{\kappa}$. The matrices $\mathbb{C}_i$ \footnote{By default, the symbol $\mathbb{C}$ or $\mathbb{D}$ would always represent a matrix with pure rational number entries in this paper.} here are composed of pure rational numbers. Formally, the solutions to Eq. (\ref{eq:dlogform}) can be expressed by a path-ordered integration,
\begin{equation}
	\vec{\text{g}}(\vec{\kappa},\epsilon) = \mathcal{P} \exp \left( \epsilon\,\int_{\gamma} 
	d \mathbb{A}(\vec{\kappa}) \right)\,\vec{\text{g}}(\vec{\kappa}_0,\epsilon),
\end{equation}
with $\vec{\text{g}}(\vec{\kappa}_0,\epsilon)$ being the values at the boundary and $\gamma$ being the path connecting $\vec{\kappa}_0$ and $\vec{\kappa}$ in kinematic variable space. In practice, if the transformation matrix doesn't involve any square roots of kinematic variables or all the roots can be rationalized simultaneously through a suitable change of variables, the solutions are straightforward to be evaluated to the Goncharov polylogarithms (GPLs) \cite{Goncharov:2001iea, Goncharov:1998kja} with a given fixed integration path (a convenient choice is to integrate along a piecewise-smooth path in kinematic space),
\begin{equation}
	G(w_n,...,w_1;a)=\int_0^a \frac{1}{t-w_n}\,G(w_{n-1},...,w_1;t)\,dt,
\end{equation}
with 
\begin{equation}
	G(w_1;a)=\int_0^a \frac{1}{t-w_1} dt ~~ w_1\neq 0, ~~~~
	G(\underbrace{0,...,0}_{n \text{ times}};a)=\frac{\log^n(a)}{n!}.
\end{equation}
The number $n$ is referred to as the \textit{weight} of a GPL and represents how many times iterated integrations are required to define the GPL. To be more specific, let $\vec{x}$ be the new variable introduced by rationalization ($\vec{x}$ remains the same as the original variable if no rationalization is needed) and suppose that all MIs are normalized to be regular at $\epsilon=0$, namely they can be expanded as a Taylor series,
\begin{equation}
	\vec{\text{g}}(\vec{x},\epsilon) = \sum_{n=0}^{\infty} \vec{\text{g}}^{(n)}(\vec{x})\,\epsilon^n,
\end{equation}
with the coefficient at each order determined order by order,
\begin{equation}
\vec{\text{g}}^{(n)}(\vec{x}) = \left\{
\begin{matrix}
	&\vec{\text{g}}^{(0)}(\vec{x}_0) &~~~~ n=0, \\
	&\vec{\text{g}}^{(n)}(\vec{x}_0)+\int_{\gamma} d\mathbb{A}(\vec{x})\vec{\text{g}}^{(n-1)}(\vec{x}) &~~~~ n>0,
\end{matrix}
\right.
\end{equation}
and $\vec{\text{g}}^{(n)}(\vec{x})$ is expressed in terms of GPLs of \textit{weight} $n$.

\input{./src/family1}
\input{./src/family2}
\input{./src/continuation}
\input{./src/checks.tex}

%% file: src/family1.tex
\subsection{Family 1}

\begin{figure}
  \centering
  \captionsetup[subfigure]{labelformat=empty}
\foreach \ii in {1,...,60}
{
\subfloat[$f_{\ii}^{\prime}$]{%
    \includegraphics[width=0.11\textwidth]{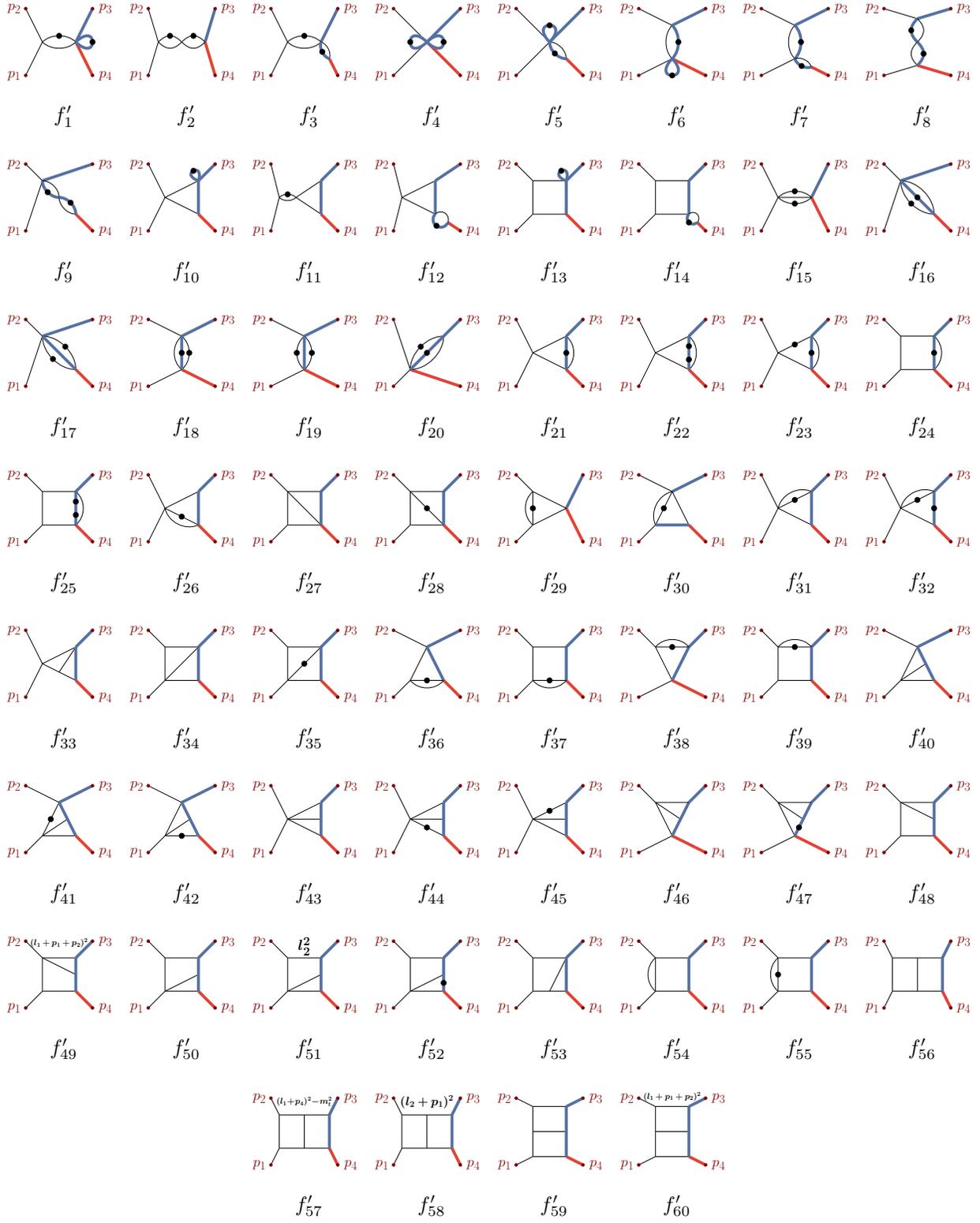}
  }
}
\caption{Master integrals of Family 1. The external momenta are displayed explicitly. Massless and massive propagators are presented by thin and thick lines, respectively. Each black dot on a line means to increase the power of the corresponding propagator by one.}
 \label{fig:mis1}
\end{figure}
Let us start with family 1. The MI vector $\vec{f}$ defined in Eq. (\ref{eq:linear1}), where MIs $f_i^{\prime}$ are displayed in Figure \ref{fig:mis1}, fulfills the differential equations w.r.t. $\vec{\kappa}$, which is linear in $\epsilon$.
\begin{align*}
f_1 = &f_1^{\prime} \,\epsilon^2 , & 
f_2 = &f_2^{\prime} \,\epsilon^2 , &
f_3 = &f_3^{\prime} \,\epsilon^2 , \quad~~&
f_4 = &f_4^{\prime} \,\epsilon^2 ,\nn
f_5 = &f_5^{\prime} \,\epsilon^2 ,&
f_6 = &f_6^{\prime} \,\epsilon^2 ,&
f_7 = &f_7^{\prime} \,\epsilon^2 ,&
f_8 = &f_8^{\prime} \,\epsilon^2 ,\nn
f_9 = &f_9^{\prime} \,\epsilon^2 ,&
f_{10} = &f_{10}^{\prime} \,\epsilon^3 ,&
f_{11} = &f_{11}^{\prime} \,\epsilon^3 ,&
f_{12} = &f_{12}^{\prime} \,\epsilon^3,\nn
f_{13} = &f_{13}^{\prime} \,\epsilon^3 ,&
f_{14} = &f_{14}^{\prime} \,\epsilon^3 ,&
f_{15} = &f_{15}^{\prime} \,\epsilon^2 ,&
f_{16} = &f_{16}^{\prime} \,\epsilon^2 ,\nn
f_{17} = &f_{17}^{\prime} \,\epsilon^2,&
f_{18} = &f_{18}^{\prime} \,\epsilon^2 ,&
f_{19} = &f_{19}^{\prime} \,\epsilon^2,&
f_{20} = &f_{20}^{\prime} \,\epsilon^2 ,\nn
f_{21} = &f_{21}^{\prime} \,\epsilon^3 ,&
f_{22} = &f_{22}^{\prime} \,\epsilon^2 ,&
f_{23} = &f_{23}^{\prime} \,\epsilon^2 ,&
f_{24} = &f_{24}^{\prime} \,\epsilon^3 ,\nn
f_{25} = &f_{25}^{\prime} \,\epsilon^2 ,&
f_{26} = &f_{26}^{\prime} \,\epsilon^3 ,&
f_{27} = &f_{27}^{\prime} \,\epsilon^4 ,&
f_{28} = &f_{28}^{\prime} \,\epsilon^3 ,\nn
f_{29} = &f_{29}^{\prime} \,\epsilon^3 ,&
f_{30} = &f_{30}^{\prime} \,\epsilon^3 ,&
f_{31} = &f_{31}^{\prime} \,\epsilon^3 ,&
f_{32} = &f_{32}^{\prime} \,\epsilon^2 ,\nn
f_{33} = &f_{33}^{\prime} \,\epsilon^4 ,&
f_{34} = &f_{34}^{\prime} \,\epsilon^4 ,&
f_{35} = &f_{35}^{\prime} \,\epsilon^3 ,&
f_{36} = &f_{36}^{\prime} \,\epsilon^3 ,\nn
f_{37} = &f_{37}^{\prime} \,\epsilon^3 ,&
f_{38} = &f_{38}^{\prime} \,\epsilon^3 ,&
f_{39} = &f_{39}^{\prime} \,\epsilon^3 ,&
f_{40} = &f_{40}^{\prime} \,\epsilon^4 ,\nn
f_{41} = &f_{41}^{\prime} \,\epsilon^3 ,&
f_{42} = &f_{42}^{\prime} \,\epsilon^3 ,&
f_{43} = &f_{43}^{\prime} \,\epsilon^4 ,&
f_{44} = &f_{44}^{\prime} \,\epsilon^3 ,\nn
f_{45} = &f_{45}^{\prime} \,\epsilon^2\,(1+2\epsilon) ,&
f_{46} = &f_{46}^{\prime} \,\epsilon^4 ,&
f_{47} = &f_{47}^{\prime} \,\epsilon^3 ,&
f_{48} = &f_{48}^{\prime} \,\epsilon^4 ,&\nn
f_{49} = &f_{49}^{\prime} \,\epsilon^4, &
f_{50} = &f_{50}^{\prime} \,\epsilon^4 ,&
f_{51} = &f_{51}^{\prime} \,\epsilon^4 ,&
f_{52} = &f_{52}^{\prime} \,\epsilon^3 ,\nn
f_{53} = &f_{53}^{\prime} \,\epsilon^4 ,&
f_{54} = &f_{54}^{\prime} \,\epsilon^3\,(1-2\epsilon ) ,&
f_{55} = &f_{55}^{\prime} \,\epsilon^3 ,&
f_{56} = &f_{56}^{\prime} \,\epsilon^4 ,\nn
f_{57} = &f_{57}^{\prime} \,\epsilon^4 ,&
f_{58} = &f_{58}^{\prime} \,\epsilon^4 ,&
f_{59} = &f_{59}^{\prime} \,\epsilon^4 ,&
f_{60} = &f_{60}^{\prime} \,\epsilon^4 .\numberthis
\label{eq:linear1}
\end{align*}
With the linear basis $\vec{f}$ at hand, by using the Magnus exponential\cite{Magnus1954OnTE,Argeri:2014qva,DiVita:2014pza} method \footnote{Alternatively, MI by MI, one can obtain the canonical basis by setting an ansatz and deriving its differential equation whose dependence on $\epsilon$ should factorize out, which in turn could fix the ansatz.}, we can construct a canonical basis as follows,
\begin{align*}
g_1 = &-s\,f_1, & 
g_2 = &s^2\,f_2,\nn
g_3 = &s\,m_w^2 f_3,&
g_4 = &f_4 ,\nn
g_5 = &-m_w^2 \,f_5  ,&
g_6 = &-t \,f_6  ,\nn
g_7 = &t \,m_w^2 \,f_7  ,&
g_8 = &t^2 \,f_8  ,\nn
g_9 = &(m_w^2)^2 \,f_9  ,&
g_{10} = &\lambda^{\prime}_1 \,f_{10}  ,\nn
g_{11} = &-s\,\lambda^{\prime}_1 \,f_{11}  ,&
g_{12} = &- m_w^2 \,\lambda^{\prime}_1 \,f_{12},\nn
g_{13} = &s \,(t-m_t^2) \,f_{13}  ,&
g_{14} = &s \,(m_t^2-t) \,m_w^2 \,f_{14}  ,\nn
g_{15} = &-s \,f_{15}  ,&
g_{16} = &-m_w^2 \,f_{16}  ,\nn
g_{17} = &(m_t^2-m_w^2) \,f_{17}  +2 m_t^2 \,f_{16}  ,&
g_{18} = &-t \,f_{18}  ,\nn
g_{19} = &(m_t^2-t) \,f_{19}  +2 m_t^2 \,f_{18}  ,&
g_{20} = &m_t^2 \,f_{20}  ,\nn
g_{21} = &\lambda^{\prime}_1 \,f_{21}  ,&
g_{22} = &\lambda^{\prime}_1 \,m_t^2 \,f_{22}  ,\nn
g_{23} = &(m_t^2-m_w^2+s)( m_t^2 \,f_{22}+\frac{3}{2} f_{21}) &
g_{24} = &s \,t \,f_{24} ,\nn
	 ~&
	-s \,m_t^2 \,f_{23},\nn
g_{25} = &-s \,m_t^2 \,(m_t^2-t) \,f_{25}  
	-s \,m_t^2 \,f_{24}  ,&
g_{26} = &\lambda^{\prime}_1 \,f_{26}  ,\nn
g_{27} = &(m_w^2-s-t) \,f_{27}  ,&
g_{28} = &s \,(t-m_t^2) \,f_{28}  ,\nn
g_{29} = &-s \,f_{29}  ,&
g_{30} = &(m_w^2-t) \,f_{30}  ,\nn
g_{31} = &\lambda^{\prime}_1 \,f_{31}  ,&
g_{32} = &
	-\frac{3}{2} (m_t^2-m_w^2+s) \,f_{31} \\
	 &~&~&
	-s \,m_t^2 \,f_{32}  
	,\nn
g_{33} = &\lambda^{\prime}_1 \,f_{33}  ,&
g_{34} = &(m_t^2-s-t) \,f_{34}  ,\nn
g_{35} = &s (t-m_t^2) \,f_{35}  ,&
g_{36} = &(m_w^2-t) \,f_{36}  ,\nn
g_{37} = &s (t-m_t^2) \,f_{37}  ,&
g_{38} = &(m_t^2-t) \,f_{38}  ,\nn
g_{39} = &s (t-m_t^2) \,f_{39}  ,&
g_{40} = &(m_w^2-t) \,f_{40}  ,\nn
g_{41} = &(t-m_t^2) (t-m_w^2) \,f_{41}  ,&
g_{42} = &(m_w^2-t) (m_w^2-m_t^2) \,f_{42}  ,\nn
g_{43} = &\lambda^{\prime}_1 \,f_{43}  ,&
g_{44} = &\lambda^{\prime}_1 \,(m_t^2-m_w^2) \,f_{44}  ,\nn
g_{45} = &\rlap{$\displaystyle 
	-s\,m_t^2\,f_{45} 
	-(m_t^2-m_w^2)^2 f_{44} +
	2s\,f_{43} 
	-s\,f_{31} 
	+\frac{1}{2} (s-3 m_t^2+3 m_w^2) f_{26} ,
	$}\nn
g_{46} = &(m_t^2-t) \,f_{46}  ,&
g_{47} = &m_t^2 \,(m_t^2-t) \,f_{47}  ,\nn
g_{48} = &\lambda^{\prime}_2 \,f_{48}  ,&
g_{49} = &
	(m_w^2-t) (f_{49}-f_{27})\\
	 &~&~&
	+s (t-m_t^2) \,f_{48} 
	,\nn
g_{50} = &\lambda^{\prime}_1 \,(m_t^2-t) \,f_{50}  ,&
g_{51} = &(m_t^2-t) (f_{51} -s \,f_{50} - f_{34})  ,\nn
g_{52} = &\rlap{$\displaystyle 
-s\,m_t^2 (m_t^2-t) f_{52} +(t-m_t^2) (m_t^2-m_w^2+s) f_{50},$} \nn
g_{53} = &s \,(t-m_w^2) \,f_{53}  ,&
g_{54} = &\lambda^{\prime}_1 \,f_{54},\nn
g_{55} = &s \,(t-m_t^2) \,f_{55}  ,&
g_{56} = &s^2 \,(m_t^2-t) \,f_{56}  ,\nn
g_{57} = &s^2 \,f_{57}  +
	s \,(m_w^2-t) (f_{55} -\frac{1}{2} f_{28})  ,&
g_{58} = &-s\,\lambda^{\prime}_1 \,f_{58}  ,\nn
g_{59} = &-s \,(m_t^2-t)^2 \,f_{59}  ,&
g_{60} = &(t-m_t^2) (t-m_w^2) \,f_{60}  ,\numberthis
\label{eq:ut1}
\end{align*}
where $\lambda^{\prime}_{1,2}$ are two square roots,
\begin{align}
\lambda^{\prime}_1 =& \sqrt{\left(s-m_t^2-m_W^2\right)^2-4\,m_t^2\,m_W^2}, \\
\lambda^{\prime}_2 =& \sqrt{s \left(t-m_t^2\right) \left(s\,t-m_t^2 \left(s+4\,t-4\,m_W^2\right)\right)}.
\end{align}
The new MIs defined in Eq. (\ref{eq:ut1}) are normalized to be finite in $\epsilon\to0$. Moreover, a simple dimensional analysis would show that they are dimensionless, so it is convenient to introduce three dimensionless variables as
\begin{equation}
	x=-\frac{s}{m_t^2}, ~~~ y=-\frac{t}{m_t^2}, ~~~ z=\frac{m_W^2}{m_t^2}.
\end{equation}
Then we can obtain the differential equations of $\vec{g}$ w.r.t. $x,y,z$,
\begin{equation}
	\partial_i\, \vec{g}(x,y,z,\epsilon) = \epsilon\, A_i(x,y,z)\,\vec{g}(x,y,z,\epsilon),
\label{eq:epsformxyz}
\end{equation}
with $A_i(x,y,z)$ containing
\begin{align}
\lambda_1 =& \sqrt{(1+x+z)^2-4z}, \\
\lambda_2 =& \sqrt{x(y+1)[x(y+1)+4(y+z)]}.
\end{align}
It is noticeable that not every single MI depends on all roots. Therefore, an optimization method is to separate the differential equation system into groups with different dependences on roots, avoiding complicating the expressions of simple MIs. According to the dependences on roots, the 60 MIs are divided into three groups, 
\begin{itemize}[align=parleft, labelsep=0em, labelwidth=6em, leftmargin=7em]
	\item [$\bullet$ no roots:] $\vec{g}_a=g_{1,...,9,15,...,20,29,30,36,38,40,41,42,46,47}$
	\item [$\bullet$ $\lambda_1$:] $\vec{g}_b=g_{1,...,7,9,10,...,45,50,...,58}$
	\item [$\bullet$ $\lambda_{1,2}$:] $\vec{g}_c=g_{4,...,8,15,...,20,26,27,28,30,31,32,34,...,52,59,60}$
\end{itemize}
The reason that there are overlaps among three groups is that for each group, the differential equation system must be complete and closed. Now we can solve the original system Eq. (\ref{eq:epsformxyz}) group by group. That is to say, once the solutions of group 1 have been fixed, the MIs of group 1 are regarded as inputs into group 2. And then, groups 1 and 2 serve as inputs into group 3.

\subsubsection{Group 1}
We can extract the differential equation system satisfied by $\vec{g}_a$ from Eq. (\ref{eq:epsformxyz}) and cast it into $d\log$ form,
\begin{equation}
	d\, \vec{g}_a(\vec{x},\epsilon) = \epsilon\, \left(\sum_{i=1}^6 
	\mathbb{C}^a_i\,d\log(\beta^a_i(\vec{x}))\right) \vec{g}_a(\vec{x},\epsilon),
\label{eq:gadlogformxyz}
\end{equation}
with $\vec{x}=(x,y,z)$ and the letters read
\begin{align}
\begin{alignedat}{3}
	\beta^a_1&=x, &\qquad
	\beta^a_2&=y, &\qquad
	\beta^a_3&=1+y, \\
	\beta^a_4&=y+z, &\qquad
	\beta^a_5&=z, &\qquad
	\beta^a_6&=1-z.
\end{alignedat}
\end{align}
They are already rational, so the solutions evaluate to GPLs quite straightforwardly. To fix the integration constants, we use the following boundary conditions:
\begin{itemize}
	\item The expressions of $g_{1,..,9,15,...,20,29,38,46,47}$ can be found in the Refs. \cite{Bonciani:2016ypc,Mastrolia:2017pfy}. So they are regarded as inputs into the system.
	\item By demanding the regularity in $y\to -z$ ($\beta^a_4\to 0$, $t\to m_W^2$), the integration constants of $g_{30,36,40,41,42}$ are fixed. 
\end{itemize}
The weights of the GPLs of arguments $x,y,z$ are each drawn from the following three sets,
\begin{equation}
	\{0\},\quad \{-1,0,-z\},\quad \{0,1\}.
\end{equation}

\subsubsection{Group 2}
For group 2, the single square root $\lambda_1$ can be rationalized by the following change of variables,
\begin{equation}
	x = (x_1-1)(1-z_1), \quad
	z = x_1 z_1.
\label{eq:xyz1}
\end{equation}
Under this transformation, we have
\begin{equation}
	\lambda_1 = x_1-z_1.
\end{equation}
Then the MIs of group 2, $\vec{g}_b$, are regarded as functions of $\vec{x}_1=(x_1,y,z_1)$, and the differential equation of $\vec{g}_b$ w.r.t. $\vec{x}_1$ fulfills the following $d\log$ form,
\begin{equation}
	d\, \vec{g}_b(\vec{x}_1,\epsilon) = \epsilon\, \left(\sum_{i=1}^{14} 
	\mathbb{C}^b_i\,d\log(\beta^b_i(\vec{x}_1))\right) \vec{g}_b(\vec{x}_1,\epsilon),
\label{eq:gbdlogformxyz}
\end{equation}
where the 14 letters are given by
\begin{align}
	\beta^b_{1}&=x_1, &
	\beta^b_{2}&=y, &
	\beta^b_{3}&=z_1, &\nn
	\beta^b_{4}&=x_1-1, &
	\beta^b_{5}&=1+y, &
	\beta^b_{6}&=1-z_1, &\nn
	\beta^b_{7}&=x_1+y, &
	\beta^b_{8}&=x_1-z_1, &
	\beta^b_{9}&=y+z_1, &\nn
	\beta^b_{10}&=1-x_1 z_1, &
	\beta^b_{11}&=y+x_1 z_1, &
	\beta^b_{12}&=x_1+y+z_1-1, &\nn
	\beta^b_{13}&=x_1 (1-z_1)+y+z_1, &
	\beta^b_{14}&
	\rlap{$\displaystyle
	=x_1+y+z_1-x_1 (y+2) z_1.
	$}
\label{eq:letter1b}
\end{align}
Again, the solutions are expressed in terms of GPLs. Note that 21 MIs of $\vec{g}_b$ have been computed in group 1. The remaining 32 integration constants are given by demanding the regularity of these MIs in some kinematic limits, which are summarized in Table \ref{tab:bound1b}. The weights of GPLs of arguments $x_1,y,z_1$ are listed in the following Table \ref{tab:weight1b}.
\begin{table}[htbp]
\renewcommand\arraystretch{2}
\begin{tabular}{cccc}
\multicolumn{3}{c}{limit} & regular MI \\
\toprule[1.5pt]
\multirow{4}*{$x_1 \to$} & $z_1$ & $(\beta^b_8\to0)$ & $g_{10,11,12,21,22,26,31,33,43,44,50,54,58}$ \\
~ & $0$ & $(\beta^b_1\to0)$ & $g_{14, 23, 32, 45, 52, 53}$ \\
~ & $-y$ & $(\beta^b_7\to0)$ & $g_{13,24,25,28,35,37,39,51,55,56,57}$ \\
~ & $1-y-z_1$ & $(\beta^b_{12}\to0)$ & $g_{27}$ \\
$z_1 \to$ & $-y$ & $(\beta^b_9\to0)$ & $g_{34}$ \\
\bottomrule[1.5pt]
\end{tabular}
\caption{The regularity conditions needed for determining unknown 32 integration constants of group 2 in family 1. The first two limits correspond to $\lambda_1\to0$ and $m_W\to0$, respectively, while the third and last one correspond to $s\to\frac{(t-m_t^2)(m_W^2-t)}{t}$ and the second last limit corresponds to $s\to m_W^2-t$.}
\label{tab:bound1b}
\end{table}
\begin{table}[htbp]
\centering
\renewcommand\arraystretch{2}
\begin{tabular}{cc}
argument & weight \\
\toprule[1.5pt]
	$x_1$ & $\left\{0,1,-y,z_1,1-y-z_1,\frac{1}{z_1},-\frac{y}{z_1},\frac{y+z_1}{z_1-1},\frac{y+z_1}{(y+2)z_1-1}\right\}$ \\
	$z_1$ & $\{0,1,-y,1-y\}$ \\
	$y$ & $\{-1,0,1\}$ \\
\bottomrule[1.5pt]
\end{tabular}
\caption{The weights of GPLs, of arguments $x_1,y,z_1$, as solutions to groups 2 of family 1.}
\label{tab:weight1b}
\end{table}

\subsubsection{Group 3}
The transformation in group 2 can not rationalize $\lambda_2$, so we need a new one to rationalize both $\lambda_1$ and $\lambda_2$,
\begin{equation}
	x= \frac{x_2}{z_2}, \quad
	y= \frac{1+x_2-x_2 z_2}{z_2+x_2 y_2-x_2y_2^2}-1, \quad
	z= \frac{(x_2+1) (z_2-1)}{z_2}.
\label{eq:xyz2}
\end{equation}
Under this new transformation, we have
\begin{equation}
\begin{split}
	\lambda_1&= \frac{x_2 z_2+1}{z_2},\\
	\lambda_2&= \frac{x_2 (2 y_2-1) (1+x_2-x_2 z_2)}{z_2 (z_2+x_2 y_2-x_2y_2^2)}.
\end{split}
\label{eq:ration2}
\end{equation}
With $\vec{x}_2=(x_2,y_2,z_2)$, the differential equation satisfied by $\vec{g}_c$ can be written as
\begin{equation}
	d\, \vec{g}_c(\vec{x}_2,\epsilon) = \epsilon\, \left(\sum_{i=1}^{18} 
	\mathbb{C}^c_i\,d\log(\beta^c_i(\vec{x}_2))\right) \vec{g}_c(\vec{x}_2,\epsilon),
\label{eq:gbdlogformxyz}
\end{equation}
where the 18 letters read
\begin{align}
	\beta^c_{1}&=x_2, &
	\beta^c_{2}&=y_2, &
	\beta^c_{3}&=z_2, &\nn
	\beta^c_{4}&=1+x_2, &
	\beta^c_{5}&=y_2-1, &
	\beta^c_{6}&=2 y_2-1, &\nn
	\beta^c_{7}&=z_2-1, &
	\beta^c_{8}&=y_2-z_2, &
	\beta^c_{9}&=y_2+z_2-1, &\nn
	\beta^c_{10}&=1+x_2 z_2, &
	\beta^c_{11}&=1+x_2 y_2, &
	\beta^c_{12}&=1-x_2 (z_2-1), &\nn
	\beta^c_{13}&=1-x_2 (y_2-1), &
	\beta^c_{14}&=z_2-x_2 y_2 (y_2-1), &\nn
	\beta^c_{15}&=z_2+y_2 (y_2-1)(1-x_2 z_2), &
	\beta^c_{16}&=
	\rlap{$\displaystyle
	z_2+x_2 z_2 (2-z_2)-x_2^2 y_2 (y_2-1),
	$} &\nn
	\beta^c_{17}&=x_2 y_2(y_2-1)+(1+x_2)(1-z_2), &
	\beta^c_{18}&=
	\rlap{$\displaystyle
	1+x_2 (2-z_2)-x_2^2 \Big[y_2(y_2-1)-(z_2-1)^2\Big].
	$}&\nn
\label{eq:letter1c}
\end{align}
Having done the computation of group 1 and 2, we are left with only $g_{48,49,59,60}$ to compute. Their expressions as functions of $\vec{x}_2$ are given by the same procedures as former groups and the integration constants are fixed in the following way:
\begin{itemize}
	\item The constants of $g_{48,49,59}$ are determined by demanding the finiteness of $g_{48,49,59}$ in the limit $\displaystyle x_2\to \frac{1}{y_2-1}$ ($\beta^c_{13}\to0$, $s\to \frac{(t-m_t^2)(m_W^2-t)}{t}$).
	\item The constants of $g_{60}$ are determined by demanding the finiteness of $g_{60}$ in the limit $\displaystyle x_2\to \frac{z_2-1}{y_2(y_2-1)-z_2+1}$ ($\beta^c_{17}\to0$, $t\to 0$).
\end{itemize}
The weights of GPLs of arguments $x_2,y_2,z_2$ are listed in the Table \ref{tab:weight1c}.
\begin{table}[htbp]
\centering
\renewcommand\arraystretch{2}
\begin{tabular}{cc}
argument & weight \\
\toprule[1.5pt]
	$x_2$ & \makecell{
		$\left\{-1,0,
		  -\frac{1}{y_2},
		  -\frac{1}{z _2},
		  \frac{1}{y_2-1},
		  \frac{1}{z_2-1},
		  \frac{z_2}{(y_2-1) y_2},
		  \frac{z_2-1}{y_2(y_2-1)-z_2+1}, \right. $ \\
		  $\left. \frac{y_2(y_2-1)+z_2}{(y_2-1) y_2 z_2},
		  \frac{(z_2-2\pm r_1)}{2[y_2(1-y_2)+(z_2-1)^2]},
		  \frac{z_2(2-z_2)\pm r_2}{2 (y_2-1) y_2}\right\}$
		}\\
	$z_2$ & $\left\{0,1,1-y_2,y_2,(1-y_2)y_2\right\}$ \\
	$y_2$ & $\left\{0,\frac{1}{2},1\right\}$ \\
\bottomrule[1.5pt]
\end{tabular}
\caption{The weights of GPLs, of arguments $x_2,y_2,z_2$, as solutions to group 3 of family 1.}
\label{tab:weight1c}
\end{table}
We used $r_{1,2}$ to represent two square roots,
\begin{equation}
\begin{split}
	r_1&=\sqrt{4y_2(y_2-1)+z_2(4-3 z_2)}, \\
	r_2&=\sqrt{z_2 [4y_2(y_2-1)+z_2(z_2-2)^2]}.
\end{split}
\end{equation}
The complete analytic expressions up to weight 4 for all MIs in family 1 are provided in the ancillary files attached to the $\texttt{arXiv}$ submission of this paper.

%% file: src/family2.tex
\subsection{Family 2}

\begin{figure}
  \centering
  \captionsetup[subfigure]{labelformat=empty}
\foreach \ii in {1,...,36}
{
\subfloat[$h_{\ii}^{\prime}$]{%
    \includegraphics[width=0.15\textwidth]{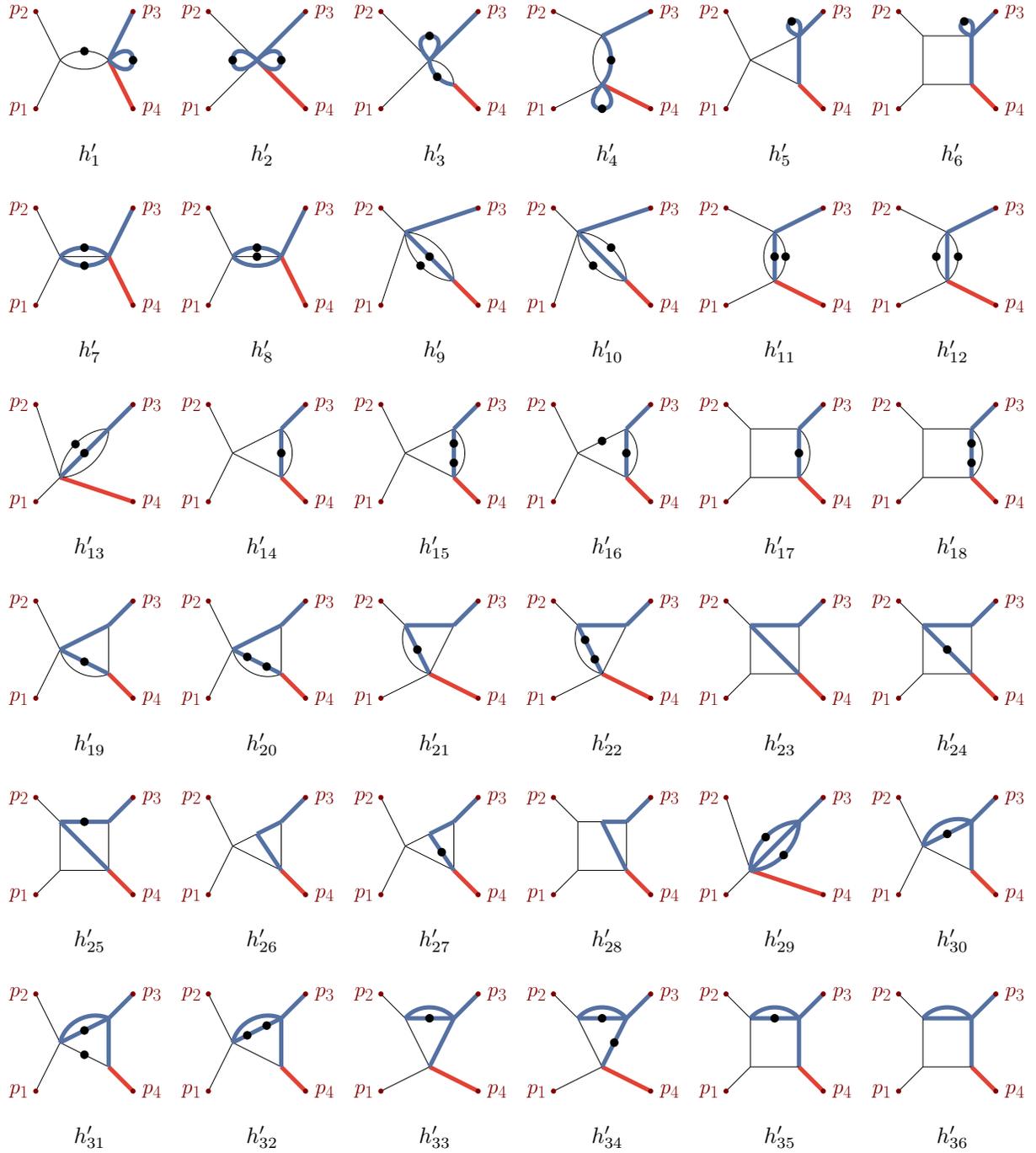}
  }
}
\caption{Master integrals of Family 2. The external momenta are displayed explicitly. Massless and massive propagators are presented by thin and thick lines respectively. Each black dot on a line means to increase the power of corresponding propagator by one.}
 \label{fig:mis2}
\end{figure}
We now turn to the computation of MIs of family 2. The chosen 36 MIs $h_i^{\prime}$ are depicted in Figure \ref{fig:mis2}. Multiplying them with factors defined in Eq. (\ref{eq:linear2}), the MI vector $\vec{h}$ fulfills a differential equation which is linear in $\epsilon$. 
\begin{align*}
h_1 = &h_1^{\prime}\,\epsilon^2, &
h_2 = &h_2^{\prime}\,\epsilon^2, & 
h_3 = &h_3^{\prime}\,\epsilon^2, & 
h_4 = &h_4^{\prime}\,\epsilon^2, \nn
h_5 = &h_5^{\prime}\,\epsilon^3, & 
h_6 = &h_6^{\prime}\,\epsilon^3, &
h_7 = &h_7^{\prime}\,\epsilon^2, & 
h_8 = &h_8^{\prime}\,\epsilon^2, \nn
h_9 = &h_9^{\prime}\,\epsilon^2, & 
h_{10} = &h_{10}^{\prime}\,\epsilon^2, &
h_{11} = &h_{11}^{\prime}\,\epsilon^2, & 
h_{12} = &h_{12}^{\prime}\,\epsilon^2, \nn
h_{13} = &h_{13}^{\prime}\,\epsilon^2, &
h_{14} = &h_{14}^{\prime}\,\epsilon^3, &
h_{15} = &h_{15}^{\prime}\,\epsilon^2, & 
h_{16} = &h_{16}^{\prime}\,\epsilon^2, \nn
h_{17} = &h_{17}^{\prime}\,\epsilon^3, &
h_{18} = &h_{18}^{\prime}\,\epsilon^2, &
h_{19} = &h_{19}^{\prime}\,\epsilon^3, & 
h_{20} = &h_{20}^{\prime}\,\epsilon^2, \nn
h_{21} = &h_{21}^{\prime}\,\epsilon^3, &
h_{22} = &h_{22}^{\prime}\,\epsilon^2, &
h_{23} = &h_{23}^{\prime}\,\epsilon^4, & 
h_{24} = &h_{24}^{\prime}\,\epsilon^3, \nn
h_{25} = &h_{25}^{\prime}\,\epsilon^3, &
h_{26} = &h_{26}^{\prime}\,\epsilon^4, &
h_{27} = &h_{27}^{\prime}\,\epsilon^2, &
h_{28} = &h_{28}^{\prime}\,\epsilon^4, \nn
h_{29} = &h_{29}^{\prime}\,\epsilon^2, &
h_{30} = &h_{30}^{\prime}\,\epsilon^3, &
h_{31} = &h_{31}^{\prime}\,\epsilon^2, &
h_{32} = &h_{32}^{\prime}\,\epsilon^2, \nn
h_{33} = &h_{33}^{\prime}\,\epsilon^3, & 
h_{34} = &h_{34}^{\prime}\,\epsilon^2, &
h_{35} = &h_{35}^{\prime}\,\epsilon^3, &
h_{36} = &h_{36}^{\prime}\,\epsilon^3\,(1-2\epsilon), \numberthis
\label{eq:linear2}
\end{align*}
With these 36 $h$ MIs, a canonical basis can be constructed as follows.
\begin{align*}
k_1 = &-s \,h_1 , &
k_2 = &h_2 , & \nn
k_3 = &-m_w^2 \,h_3 , & 
k_4 = &-t \,h_4 , & \nn
k_5 = &\lambda^{\prime}_1 \,h_5 , & 
k_6 = &s \,(t-m_t^2) \,h_6 , & \nn
k_7 = &-s \,h_7 , & 
k_8 = &\lambda^{\prime}_3 \,h_8  +\frac{1}{2} \lambda^{\prime}_3 \,h_7 , & \nn
k_9 = &-m_w^2 \,h_9 , & 
k_{10} = &(m_t^2-m_w^2) \,h_{10}  +2 m_t^2 \,h_9 , & \nn
k_{11} = &-t \,h_{11} , & 
k_{12} = &(m_t^2-t) \,h_{12}  +2 m_t^2 \,h_{11} , & \nn
k_{13} = &m_t^2 \,h_{13} , &
k_{14} = &\lambda^{\prime}_1 \,h_{14} , & \nn
k_{15} = &\lambda^{\prime}_1 \,m_t^2 \,h_{15} , & 
k_{16} = &(m_t^2-m_w^2+s) (m_t^2 h_{15} +\frac{3}{2} h_{14})\\
	 &~&~&
	 -s \,m_t^2 \,h_{16}, & \nn
k_{17} = &s \,t \,h_{17} , &
k_{18} = &-s \,m_t^2 \,(m_t^2-t) \,h_{18}  -s \,m_t^2 \,h_{17} , & \nn
k_{19} = &\lambda^{\prime}_1 \,h_{19} , & 
k_{20} = &\lambda^{\prime}_1 \,m_t^2\, h_{20} , & \nn
k_{21} = &(m_t^2-t) \,h_{21} , &
k_{22} = &m_t^2 \,(m_t^2-t) \,h_{22} , & \nn
k_{23} = &(m_w^2-s-t) \,h_{23} , & 
k_{24} = &\lambda^{\prime}_2 \,h_{24} , & \nn
k_{25} = &\frac{1}{2} [m_t^2 \,(2 m_w^2-s-2 t)+s \,t] \,h_{24} &
k_{26} = &\lambda^{\prime}_1 \,h_{26} , \nn
	 ~&
	 +m_t^2 \,(m_w^2-t) \,h_{25}, &\nn
k_{27} = &
\rlap{$\displaystyle
	-s \,m_t^2 \,h_{27}  
	-2 (m_t^2-m_w^2+s) \,h_{26}  
	+2 m_t^2 \,(m_t^2-m_w^2) \,h_{20}  
	+(m_t^2-m_w^2+s) \,h_{19}  
	$}& \nn
\rlap{$\displaystyle
	+2 s \,m_t^2 \,h_{15}  
	+(m_t^2-m_w^2+s) \,h_{14},
	$}& \nn
k_{28} = &s \,(t-m_t^2) \,h_{28} , & 
k_{29} = &m_t^2 \,h_{29} , & \nn
k_{30} = &\lambda^{\prime}_1 \,h_{30} , & 
k_{31} = &\lambda^{\prime}_1 \,(m_t^2-m_w^2) \,h_{31} +\lambda^{\prime}_1 \,h_8 \\
	 &~&~&
	+\frac{1}{2} \lambda^{\prime}_1 \,h_7 , & \nn
k_{32} = &
\rlap{$\displaystyle
	-\frac{s \,(2 m_t^2+2 m_w^2-s)}{m_t^2+m_w^2-s}(\,m_t^2 \,h_{32}+\frac{3}{2}\,h_{30})
	+\frac{(m_t^2-m_w^2)^2 (m_t^2+m_w^2)}{2 (m_t^2+m_w^2-s)} \,h_{31}  
	$}& \nn
\rlap{$\displaystyle
	-\frac{3 s \,m_t^2}{2 (m_t^2+m_w^2-s)} \,h_{29}  
	+\frac{[(m_t^2)^2-(m_w^2)^2]}{2 (m_t^2+m_w^2-s)}(\,h_8+\frac{1}{2}\,h_7),
	$}& \nn
k_{33} = &(m_t^2-t) \,h_{33} , & 
k_{34} = &(m_t^2-t) (m_t^2+t) \,h_{34}  +\frac{3 t }{2} \,h_{29}, & \nn
k_{35} = &\lambda^{\prime}_2 \,h_{35} , &
k_{36} = &[m_t^2 (s+4 t-4 m_w^2)-s \,t] \,h_{35} \\
	 &~&~&
	 +(m_w^2-t) \,h_{36}, \numberthis
\label{eq:ut2}
\end{align*}
where a new square root occurs in the definition of $k_{7,8}$,
\begin{equation}
\lambda^{\prime}_3 = \sqrt{s\left(s-4m_t^2\right)}.
\end{equation}
The canonical basis in Eq. (\ref{eq:ut2}) is normalized to be dimensionless as well and again, its differential equation system in $\vec{x}$ can be constructed like family 1 (Eq. (\ref{eq:epsformxyz})) but involves three square roots simultaneously,
\begin{align}
\lambda_1 =& \sqrt{(1+x+z)^2-4z}, \\
\lambda_2 =& \sqrt{x(y+1)[x(y+1)+4(y+z)]}, \\
\lambda_3 =& \sqrt{x(x+4)}.
\end{align}
Following the same procedures as family 1, we separate the 36 MIs into three groups \footnote{We do not identify MIs that only depend on $\lambda_1$ or $\lambda_3$ since they have been computed in family 1 or available in the literature.},
\begin{itemize}[align=parleft, labelsep=0em, labelwidth=6em, leftmargin=7em]
	\item [$\bullet$ no roots:] $\vec{k}_a=k_{1,...,4,9,...,13,21,22,29,33,34}$
	\item [$\bullet$ $\lambda_{1,3}$:] $\vec{k}_b=k_{1,...,20,26,27,29,...,32}$
	\item [$\bullet$ $\lambda_{1,2,3}$:] $\vec{k}_c=\vec{k}_{1,...,4,7,...,36}$
\end{itemize}
The two families share some MIs, namely $k_{1,...,6,9,...,18}$ in family 2 correspond to $g_{1,4,5,6,10,13,16,...,25}$ in family 1. In addition, the expressions of $k_{7,8,21,22,29,33,34}$ can be found in Ref. \cite{Mastrolia:2017pfy}. Therefore the MIs of group 1 are all known, and we only need to compute MIs in groups 2 and 3, more precisely, $k_{19,20,26,27,30,31,32}$ in group 2 and $k_{23,24,25,28,35,36}$ in group 3.

\subsubsection{Group 2}
By applying the change of variables as
\begin{equation}
\begin{split}
	x &=\frac{(1-x_3)^2}{x_3}, \\
	z &=\frac{(1-x_3+z_3)(x_3-z_3+x_3z_3)}{x_3 z_3},
\end{split}
\label{eq:xyz3a}
\end{equation}
$\lambda_1$ and $\lambda_3$ are rationalized at the same time,
\begin{equation}
\begin{split}
	\lambda_1 &= \frac{(1-x_3)(x_3+z_3^2)}{x_3 z_3}, \\
	\lambda_3 &= \frac{(1-x_3)(1+x_3)}{x_3}.
\end{split}
\end{equation}
With $\vec{x}^{\,\prime}_3=(x_3,y,z_3)$, the differential equation of $\vec{k}_b$ w.r.t. $\vec{x}^{\,\prime}_3$ fulfills the following $d\log$ form,
\begin{equation}
	d\, \vec{k}_b(\vec{x}^{\,\prime}_3,\epsilon) = \epsilon\, \left(\sum_{i=1}^{16} 
	\mathbb{D}^b_i\,d\log(\eta^b_i(\vec{x}^{\,\prime}_3))\right) \vec{k}_b(\vec{x}^{\,\prime}_3,\epsilon),
\end{equation}
where the 16 letters are given by
\begin{align}
	\eta^b_{1}&=x_3, &
	\eta^b_{2}&=y, &
	\eta^b_{3}&=z_3, &\nn
	\eta^b_{4}&=1+x_3, &
	\eta^b_{5}&=1-x_3, &
	\eta^b_{6}&=1+y, &\nn
	\eta^b_{7}&=1+z_3, &
	\eta^b_{8}&=z_3-x_3, &
	\eta^b_{9}&=x_3+z_3^2, &\nn
	\eta^b_{10}&=1-x_3+z_3, &
	\eta^b_{11}&=1-x_3+2 z_3, &
	\eta^b_{12}&=1-x_3+z_3+y z_3, &\nn
	\eta^b_{13}&=x_3 (z_3+1)-z_3, &
	\eta^b_{14}&=x_3 (z_3+2)-z_3, &
	\eta^b_{15}&=x_3-z_3+y x_3+x_3 z_3, &\nn
	\eta^b_{16}&=
	\rlap{$\displaystyle
	x_3 y z_3+(1-x_3+z_3) (x_3-z_3+x_3 z_3).
	$}
\label{eq:letter2b}
\end{align}
The integration leads to results in terms of GPLs. In order to fix the integration constants, the following conditions are used:
\begin{itemize}
	\item The constants of $k_{19,20,26,30,31}$ are determined by the finiteness of these MIs in the limit $\displaystyle x_3\to-z_3^2$ ($\eta^b_{9}\to0$, $\lambda_1\to0$).
	\item The constants of $g_{32}$ are determined by its regularity in the limit $\displaystyle x_3\to \frac{z_3}{z_3+1}$ ($\eta^b_{14}\to0$, $m_W\to0$).
	\item The constants of $g_{27}$ are determined by matching against its counterpart in the case $m_W=m_t$ provided by Ref. \cite{Becchetti:2019tjy}.
\end{itemize}
The weights of GPLs of arguments $x_3,y,z_3$ are listed in the Table \ref{tab:weight2b}.
\begin{table}[htbp]
\centering
\renewcommand\arraystretch{2}
\begin{tabular}{cc}
argument & weight \\
\toprule[1.5pt]
	$y$ & $\left\{-1,0,\frac{x_3-z_3-1}{z_3},
	   \frac{z_3-x_3-x_3z_3}{x_3},\frac{(x_3-z_3-1)(x_3-z_3+x_3 z_3)}{x_3 z_3}\right\}$ \\
	$x_3$ & $\left\{-1,0,1,z_3,-z_3^2,z_3+1,2 z_3+1,\frac{z_3}{z_3+1},\frac{z_3}{z_3+2}\right\}$ \\
	$z_3$ & $\left\{-1,-\frac{1}{2},0\right\}$ \\
\bottomrule[1.5pt]
\end{tabular}
\caption{The weights of GPLs, of arguments $x_3,y,z_3$, as solutions to group 2 of family 2.}
\label{tab:weight2b}
\end{table}

\subsubsection{Group 3}
Together with Eq. (\ref{eq:xyz3a}), adding the following new transformation,
\begin{equation}
	y=\frac{(1-x_3) (z_3+1) (z_3-x_3)}{z_3(1-y_3) (x_3+y_3)}-1,
\label{eq:xyz3}
\end{equation}
can rationalize all 3 square roots at the same time,
\begin{equation}
\begin{split}
	\lambda_1 &= \frac{(1-x_3)(x_3+z_3^2)}{x_3 z_3}, \\
	\lambda_2 &= \frac{(1-x_3){}^2 (z_3+1) (x_3+2 y_3-1)(x_3-z_3)}{x_3 z_3(1-y_3) (x_3+y_3)}, \\
	\lambda_3 &= \frac{(1-x_3)(1+x_3)}{x_3}.
\end{split}
\label{eq:ration2}
\end{equation}
Then the differential equation satisfied by $\vec{k}_c$ can be written as
\begin{equation}
	d\, \vec{k}_c(\vec{x}_3,\epsilon) = \epsilon\, \left(\sum_{i=1}^{23} 
	\mathbb{D}^c_i\,d\log(\eta^c_i(\vec{x}_3))\right) \vec{k}_c(\vec{x}_3,\epsilon),
\label{eq:gbdlogformxyz}
\end{equation}
where $\vec{x}_3=(x_3,y_3,z_3)$ and we have more letters,
\begin{align}
	\eta^c_{1}&=x_3, &
	\eta^c_{2}&=y_3, &
	\eta^c_{3}&=z_3, &\nn
	\eta^c_{4}&=1+x_3, &
	\eta^c_{5}&=1-x_3, &
	\eta^c_{6}&=1+z_3, &\nn
	\eta^c_{7}&=x_3+y_3, &
	\eta^c_{8}&=1-y_3, &
	\eta^c_{9}&=y_3+z_3, &\nn
	\eta^c_{10}&=z_3-x_3, &
	\eta^c_{11}&=x_3+z_3^2, &
	\eta^c_{12}&=y_3 z_3-x_3, &\nn
	\eta^c_{13}&=x_3+y_3-1, &
	\eta^c_{14}&=x_3+2 y_3-1, &
	\eta^c_{15}&=1-x_3+z_3, &\nn
	\eta^c_{16}&=1-x_3+2 z_3, &
	\eta^c_{17}&=1-x_3-y_3+z_3, &
	\eta^c_{18}&=x_3 (z_3+1)-z_3, &\nn
	\eta^c_{19}&=x_3 (z_3+2)-z_3, &
	\eta^c_{20}&=x_3-z_3+x_3 z_3+y_3 z_3, &\nn
	\eta^c_{21}&=
	\rlap{$\displaystyle
	(x_3-z_3-1) (x_3-z_3+x_3 z_3)+(x_3-1) y_3 z_3+y_3^2 z_3,
	$} &\nn
	\eta^c_{22}&=
	\rlap{$\displaystyle
	(1-x_3) x_3 z_3+(1-x_3) y_3 (x_3-z_3^2)-y_3^2 (x_3-z_3^2),
	$} &\nn
	\eta^c_{23}&=
	\rlap{$\displaystyle
	(x_3-z_3-1) (x_3-z_3)+x_3 z_3 (x_3-z_3-2)+2 (x_3-1) y_3 z_3+2 y_3^2 z_3.
	$}
\end{align}
For this group, the integration constants are fixed by using the regularity conditions as follows.
\begin{itemize}
	\item Being regular in the limit $y_3\to \tau_1$ ($t\to m_W^2-s$) fixes the constants of $k_{23}$. Here, $\tau_1$ stands for the zero of $\eta^c_{22}=0$.
	\item Being regular in the limit $\displaystyle y_3\to -z_3$ ($\eta^c_{9}\to0$, $s\to \frac{(t-m_t^2)(m_W^2-t)}{t}$) fixes the constants of $k_{24,25,28,35,36}$.
\end{itemize}
Due to the complexity of these letters, the weights of relevant GPLs are more complicated than those in previous groups. They are given in Table \ref{tab:weight2c},
\begin{table}[htbp]
\centering
\renewcommand\arraystretch{2}
\begin{tabular}{cc}
argument & weight \\
\toprule[1.5pt]
	$y_3$ & \makecell{
		$\left\{0,1,
		  -x_3,
		  -z_3,
		  1-x_3,
		  \frac{1-x_3}{2},
		  \frac{x_3}{z_3},
		  1-x_3+z_3, \right. $ \\
		  $ \left.
		  \frac{z_3-x_3-x_3z_3}{z_3},
		  \frac{1-x_3\pm r_4}{2},
		  \frac{1-x_3\pm r_5}{2},
		  \frac{1-x_3\pm r_6}{2}\right\}$
		}\\
	$x_3$ & \makecell{
		$\left\{-1,0,1,z_3,-z_3^2,
		  z_3+1,
		  2 z_3+1, \right. $ \\
		  $ \left.
	          \frac{z_3}{z_3+1},
		  \frac{z_3}{z_3+2},
		  \frac{z_3^2+4z_3+1\pm r_3}{2(z_3+1)}\right\}$
		  }\\
	$z_3$ & $\left\{-1,-\frac{1}{2},0\right\}$ \\
\bottomrule[1.5pt]
\end{tabular}
\caption{The weights of GPLs, of arguments $x_3,y_3,z_3$, as solutions to group 3 of family 2.}
\label{tab:weight2c}
\end{table}
where another four square roots come into play,
\begin{equation}
\begin{split}
	r_3&=\sqrt{(z_3+1){}^4+4 z_3^2}, \\
	r_4&=\sqrt{\frac{(1-x_3+2 z_3) (2x_3-z_3+x_3z_3)}{z_3}},\\
	r_5&=\sqrt{\frac{(x_3-1) [(x_3-z_3)^2-x_3 (z_3+1)^2]}{x_3-z_3^2}},\\
	r_6&=\sqrt{(1-x_3)^2+\frac{4 (1-x_3+z_3) (x_3-z_3+x_3z_3)}{z_3}}.
\end{split}
\end{equation}
The complete analytic expressions up to weight 4 for all MIs in family 2 are provided in the ancillary files attached to the $\texttt{arXiv}$ submission of this paper.

%% file: src/continuation.tex
\subsection{Analytic Continuation}

Usually, the integration of the differential equation system is performed in one piece of whole kinematic space. With the obtained expressions, their evaluations in other regions can be achieved through analytic continuation. The chosen region is always determined by requiring to satisfy some properties, such as below the production threshold, and accurate values can be provided by a numerical routine efficiently. For our problem, we first perform the calculation in the region
\begin{equation}
	s<0 \wedge t<0 \wedge 0<m_W<m_t.
\label{eq:nonphyreg}
\end{equation}
In comparison, the region for on-shell $t$ and $W$ production fulfills the following kinematic constraints,
\begin{equation}
	s\geqslant(m_t+m_W)^2 \wedge 0<m_W<m_t \wedge
	\frac{m_t^2+m_W^2-s-\lambda_1^{\prime}}{2}\leqslant t \leqslant \frac{m_t^2+m_W^2-s+\lambda_1^{\prime}}{2} < 0,
\label{eq:phyreg}
\end{equation}
where, in the center-of-mass frame of colliding partons, the lower and upper boundaries of $t$ correspond to the case when outgoing particles fly parallelly to the beam line. The region above translates to
\begin{equation}
	-x\geqslant(1+\sqrt{z})^2 \wedge 0<z<1 \wedge
	0<\frac{-1-z-x-\lambda_1}{2}\leqslant y \leqslant \frac{-1-z-x+\lambda_1}{2}.
\end{equation}
With a given specific point in kinematic space, we need to invert the transformations that rationalize the square roots to express the new variables in terms of $x,y,z$. Due to the nonlinearity of the transformations, it is possible to derive multiple inverse solutions. Our choices are summarized as follows.
\begin{itemize}
	\item
	Group 2 of family 1 (involving $\lambda_1$)
\begin{equation}
\begin{split}
	x_1&=\frac{x+z+1+\lambda_1}{2}, \\
	z_1&=\frac{x+z+1-\lambda_1}{2}.
\end{split}
\label{eq:solf1g2}
\end{equation}
	\item
	Group 3 of family 1 (involving $\lambda_{1,2}$)
\begin{equation}
\begin{split}
	x_2&=\frac{x+z-1+\lambda_1}{2}, \\
	y_2&=\frac{x(y+1)+\lambda_2}{2x(y+1)}, \\
	z_2&=\frac{x+z-1+\lambda_1}{2x}.
\end{split}
\label{eq:solf1g3}
\end{equation}
	\item
	Group 2 and 3 of family 2 (involving $\lambda_{1,2,3}$)
\begin{equation}
\begin{split}
	x_3&=\frac{\lambda_3-x}{\lambda_3+x}, \\
	y_3&=\frac{x_3}{x_3-1}\frac{x(y+1)+\lambda_2}{2(y+1)}, \\
	z_3&=\frac{x_3}{1-x_3}\frac{x+z-1-\lambda_1}{2}.
\end{split}
\label{eq:solf2g23}
\end{equation}
\end{itemize}
Generally, one has to be careful when assigning values to $x,y,z$. It should be kept in mind that they each carry a positive or negative vanishing imaginary part that originates from the Mandelstam variables and masses, i.e. the Feynman prescription,
\begin{equation}
	s+i0^+,~~ t+i0^+,~~ m_W^2+i0^+,
\end{equation}
while keeping $m_t^2$ real. That is to say
\begin{equation}
	x-i0^+,~~ y-i0^+,~~ z+i0^+.
\end{equation}
The reason for this is to make sure that the GPLs are evaluated in the right branch since they are, generally, multi-valued functions. However, it is safe to carry the \textit{wrong} imaginary parts if no ambiguity occurs between positive and negative imaginary parts, i.e. not crossing branch cut. In the region defined by Eq. (\ref{eq:nonphyreg}), going through Eq. (\ref{eq:solf1g2}-\ref{eq:solf2g23}) without considering the imaginary parts would give the correct values for all MIs. When $x,y,z$ live in the space according to Eq. (\ref{eq:phyreg}), the imaginary parts matter. Practically, one can assign small finite imaginary parts to $x,y,z$ and compute $x_i,y_i,z_i$ according to Eq. (\ref{eq:solf1g2}-\ref{eq:solf2g23}). However, that makes the expressions quite cumbersome. We instead first obtain the values with real inputs, $x_i^{\prime},y_i^{\prime},z_i^{\prime}$ and later let them carry imaginary parts that would reproduce correct imaginary parts of $x,y,z$. In the current region, we further distinguish two cases, $s<4m_t^2$ and $s>4m_t^2$. To summarize, the appropriate choices for $x_i,y_i,z_i$ in different regions are listed in Table \ref{tab:xiyizi}.
\begin{table}[htbp]
\centering
\renewcommand\arraystretch{2}
\begin{tabular}{ccC{1cm}C{1cm}cccC{1cm}C{1cm}cC{1cm}C{1cm}}
region& $x$ & $y$ & $z$ & $x_1$ & $z_1$ & $x_2$ & $y_2$ & $z_2$ & $x_3$ & $y_3$ & $z_3$ \\
\toprule[1.5pt]
$s<0$& $x$ & $y$ & $z$ & $x_1^{\prime}$ & $z_1^{\prime}$ & $x_2^{\prime}$ & $y_2^{\prime}$ & $z_2^{\prime}$ & $x_3^{\prime}$ & $y_3^{\prime}$ & $z_3^{\prime}$ \\
$s<4m_t^2$ & 
	\multirow{2}*{$x-i0^+$} & 
	\multirow{2}*{$y$} & 
	\multirow{2}*{$z$} & 
	\multirow{2}*{$x_1^{\prime}+i0^+$} & 
	\multirow{2}*{$z_1^{\prime}-i0^+$} & 
	\multirow{2}*{$x_2^{\prime}-i0^+$} & 
	\multirow{2}*{$y_2^{\prime}$} & 
	\multirow{2}*{$z_2^{\prime}$} & 
	$e^{i\phi}$ & 
	$y_3^{\prime}$ & 
	$z_3^{\prime}$ \\
$s>4m_t^2$ & 
	~ & 
	~ & 
	~ & 
	~ & 
	~ & 
	~ & 
	~ & 
	~ & 
	$x_3^{\prime}+i0^+$ & 
	$y_3^{\prime}$ & 
	$z_3^{\prime}$ \\
\end{tabular}
	\caption{The solutions for $x_i,y_i,z_i$ with explicit imaginary parts in different regions of kinematic space. We have used $\phi=2\,\text{arctan}(\sqrt{-x/(x+4)})$ and assumed that $s>(m_t+m_W)^2$ when $s>0$. Also, note that $t$ and $m_W$ should satisfy the constraints according to Eq. (\ref{eq:nonphyreg}) and (\ref{eq:phyreg}) for $s<0$ and $s>0$, respectively.} 
\label{tab:xiyizi}
\end{table}

Caution should be exercised at the threshold, i.e. $s=4m_t^2$ or $s=(m_t+m_W)^2$, since we derive the expressions off the threshold. In principle, it is not safe to approach these two limits starting from our general solutions because divergences might show up. One of the correct ways is to take the limit at the very beginning, before the IBP reduction, and then compute the found MIs. Fortunately, it is safe to use our expressions directly for $s=(m_t+m_W)^2$. The corresponding values of $x_i,y_i,z_i$ are the same as $s<4m_t^2$. However, the evaluations of the MIs at $s=4m_t^2$ deserve extra investigations which are not included in this paper.

%% file: src/checks.tex
\subsection{Numerical Checks}

The validities of our analytic expressions can be checked by comparing them against their counterparts provided by an independent method such as the sector decomposition. After choosing some points in kinematic space, great agreement is found between our analytic results and numerical evaluations from \texttt{pySecDec} \cite{Borowka:2017idc,Borowka:2018goh} except for the complicated MIs like $g_{48,...,53,56,...,60}$. When dealing with GPLs, we have used the package \texttt{PolyLogTools} \cite{Maitre:2005uu,Maitre:2007kp,Duhr:2019tlz} and library \texttt{GiNaC} \cite{Bauer:2000cp,Vollinga:2004sn}. It is quite challenging for a numerical routine to obtain accurate values for these six- and seven-line MIs, especially in the region defined by Eq. (\ref{eq:phyreg}). In order to obtain high-precision values for all MIs, we first compute them in the limit $s\to 0, t\to 0, m_W\to 0$ and $m_t\to 1$. The typical behavior of a canonical MI, say $g$, in this limit is
\begin{equation}
	g \sim \sum c_i(\epsilon)\,x^{-\sigma_i \epsilon}, ~~~ \sigma_i \in \{0,1,2,3,4\}.
\end{equation}
What we need to do is to compute the coefficients $c_i(\epsilon)$ up to the relevant order in $\epsilon$. One way for this is to use the expansion by regions strategy with the public code \texttt{asy.m} \cite{Jantzen:2012mw,Pak:2010pt} (which is now part of the program \texttt{FIESTA} \cite{Smirnov:2015mct,Smirnov:2021rhf}). Then the remaining integration of Feynman parameters is performed by utilizing the \texttt{HyperInt} package \cite{Panzer:2014caa}. It turns out that all the coefficients can be fixed except for the MIs $g_{56,57,58,59}$ in the region where they behave like $x^{-2\epsilon}$. Too many computer memory resources are required to compute these coefficients. We instead use the Mellin-Barnes(MB) representation method \cite{Smirnov:1999gc,Tausk:1999vh} to compute the remaining $c_i(\epsilon)$. More precisely, the MB representations of the MIs are obtained by using the package \texttt{AMBRE} \cite{Gluza:2010rn} and \texttt{MB} \cite{Czakon:2005rk}. And then, \texttt{MBasymptotic} \cite{Czakon:mbasy} provides their asymptotic expansions, which are transformed into up to fold-2 MB integrals by applying the Barnes lemmas \cite{Kosower:barnes}. To the end, \texttt{MBsums} \cite{Ochman:2015fho} helps to do the residue calculations and expresses the results in terms of multiple sums whose evaluations are performed with the packages \texttt{Sigma} \cite{schneider2007symbolic}, \texttt{EvaluteMultiSums} \cite{Schneider2013,Schneider:2013zna} and \texttt{HarmonicSums} \cite{Ablinger:2009ovq,Ablinger:2011te,Ablinger:2013cf,Ablinger:2012ufz}. Finally, we obtain 
\begin{align*}
g_1&= x^{-\epsilon} \left(1-\zeta_2 \epsilon^2-2 \zeta_3 \epsilon^3-\frac{9 \zeta_4}{4} \epsilon^4\right)+\mathcal{O}(\epsilon^5), \nn
g_2&= x^{-2\epsilon} \left(1-2 \zeta_2 \epsilon^2-4 \zeta_3 \epsilon^3-2 \zeta_4 \epsilon^4\right)+\mathcal{O}(\epsilon^5), \nn
g_4&= 1, \nn
g_{10}&\sim -\frac{1}{2}+x^{-\epsilon} \left(1-\zeta_2 \epsilon^2-2 \zeta_3 \epsilon^3-\frac{9 \zeta_4}{4} \epsilon^4\right)+x^{-2\epsilon} \left(-\frac{1}{2}-2 \zeta_2 \epsilon^2-14 \zeta_4 \epsilon^4\right)+\mathcal{O}(\epsilon^5), \nn
g_{11}&\sim 
\rlap{$\displaystyle
x^{-\epsilon} \left(-\frac{1}{2}+\frac{\zeta_2}{2} \epsilon^2+\zeta_3 \epsilon^3+\frac{9 \zeta_4}{8} \epsilon^4\right)
+x^{-2\epsilon} \left(1-2 \zeta_2 \epsilon^2-4 \zeta_3 \epsilon^3-2 \zeta_4 \epsilon^4\right)
$}  \nn
\rlap{$\displaystyle\hspace{1cm}
+x^{-3\epsilon} \left(-\frac{1}{2}-\frac{3 \zeta_2}{2} \epsilon^2+\zeta_3 \epsilon^3-\frac{63 \zeta_4}{8} \epsilon^4\right)+\mathcal{O}(\epsilon^5), 
$}  \nn
g_{13}&\sim x^{-\epsilon} \left(2-2 \zeta_2 \epsilon^2-4 \zeta_3 \epsilon^3-\frac{9 \zeta_4}{2} \epsilon^4\right)
+x^{-2\epsilon} \left(-\frac{1}{2}-2 \zeta_2 \epsilon^2-14 \zeta_4 \epsilon^4\right)+\mathcal{O}(\epsilon^5), \nn
g_{15}&= x^{-2\epsilon} \left(-1+2 \zeta_2 \epsilon^2+10 \zeta_3 \epsilon^3+11 \zeta_4 \epsilon^4\right)+\mathcal{O}(\epsilon^5), \nn
g_{17}&= -1-2 \zeta_2 \epsilon^2+2 \zeta_3 \epsilon^3-9 \zeta_4 \epsilon^4+\mathcal{O}(\epsilon^5), \nn
g_{19}&= -1-2 \zeta_2 \epsilon^2+2 \zeta_3 \epsilon^3-9 \zeta_4 \epsilon^4+\mathcal{O}(\epsilon^5), \nn
g_{20}&\sim -\frac{1}{4}-\zeta_2 \epsilon^2-2 \zeta_3 \epsilon^3-16 \zeta_4 \epsilon^4+\mathcal{O}(\epsilon^5), \nn
g_{21}&\sim \left(\zeta_2 \epsilon^2+5 \zeta_3 \epsilon^3+\frac{55 \zeta_4}{2} \epsilon^4\right)
+x^{-\epsilon} \left(-\zeta_2 \epsilon^2-6 \zeta_3 \epsilon^3-\frac{81 \zeta_4}{4} \epsilon^4\right)+\mathcal{O}(\epsilon^5), \nn
g_{22}&\sim 
\rlap{$\displaystyle
\left(\frac{1}{8}-\frac{\zeta_2}{2} \epsilon^2-4 \zeta_3 \epsilon^3-\frac{39 \zeta_4}{2} \epsilon^4\right)
+x^{-\epsilon} \left(-\frac{1}{6}+\frac{7 \zeta_2}{6} \epsilon^2+\frac{19 \zeta_3}{3} \epsilon^3+\frac{165 \zeta_4}{8} \epsilon^4\right)
$}  \nn
\rlap{$\displaystyle\hspace{1cm}
+x^{-4\epsilon} \left(\frac{1}{24}+\frac{5 \zeta_2}{6} \epsilon^2-\frac{\zeta_3}{3} \epsilon^3+27 \zeta_4 \epsilon^4\right)+\mathcal{O}(\epsilon^5), 
$}  \nn
g_{23}&\sim 
\rlap{$\displaystyle
\left(\frac{1}{8}+\zeta_2 \epsilon^2+\frac{7 \zeta_3}{2} \epsilon^3+\frac{87 \zeta_4}{4} \epsilon^4\right)
+x^{-\epsilon} \left(\frac{1}{6}+\frac{\zeta_2}{3} \epsilon^2+\frac{8 \zeta_3}{3} \epsilon^3+\frac{39 \zeta_4}{4} \epsilon^4\right)
$}  \nn
\rlap{$\displaystyle\hspace{1cm}
+x^{-4\epsilon} \left(-\frac{1}{24}-\frac{5 \zeta_2}{6} \epsilon^2+\frac{\zeta_3}{3} \epsilon^3-27 \zeta_4 \epsilon^4\right)+\mathcal{O}(\epsilon^5), 
$}  \nn
g_{25}&\sim x^{-\epsilon} \left(-\frac{2}{3}+\frac{5 \zeta_2}{3} \epsilon^2+\frac{22 \zeta_3}{3} \epsilon^3+\frac{87 \zeta_4}{4} \epsilon^4\right)
+x^{-4\epsilon} \left(\frac{1}{24}+\frac{5 \zeta_2}{6} \epsilon^2-\frac{\zeta_3}{3} \epsilon^3+27 \zeta_4 \epsilon^4\right)+\mathcal{O}(\epsilon^5), \nn
g_{26}&\sim 
\rlap{$\displaystyle
\left(\frac{1}{6}+\frac{\zeta_2}{3} \epsilon^2-\frac{\zeta_3}{3} \epsilon^3+\frac{3 \zeta_4}{2} \epsilon^4\right)
+x^{-2\epsilon} \left(-\frac{1}{2}+\zeta_2 \epsilon^2+5 \zeta_3 \epsilon^3+\frac{11 \zeta_4}{2} \epsilon^4\right)
$}  \nn
\rlap{$\displaystyle\hspace{1cm}
+x^{-3\epsilon} \left(\frac{1}{3}+\frac{5 \zeta_2}{3} \epsilon^2-\frac{8 \zeta_3}{3} \epsilon^3+\frac{69 \zeta_4}{4} \epsilon^4\right)+\mathcal{O}(\epsilon^5), 
$}  \nn
g_{28}&\sim x^{-2\epsilon} \left(-\frac{3}{2}+3 \zeta_2 \epsilon^2+15 \zeta_3 \epsilon^3+\frac{33 \zeta_4}{2} \epsilon^4\right)
+x^{-3\epsilon} \left(\frac{2}{3}+\frac{10 \zeta_2}{3} \epsilon^2-\frac{16 \zeta_3}{3} \epsilon^3+\frac{69 \zeta_4}{2} \epsilon^4\right)+\mathcal{O}(\epsilon^5), \nn
g_{29}&\sim x^{-2\epsilon} \left(\frac{1}{4}-2 \zeta_3 \epsilon^3-3 \zeta_4 \epsilon^4\right)+\mathcal{O}(\epsilon^5), \nn
g_{31}&\sim 
\rlap{$\displaystyle
\left(\frac{1}{8}+\frac{\zeta_2}{2} \epsilon^2+\zeta_3 \epsilon^3+8 \zeta_4 \epsilon^4\right)
+x^{-2\epsilon} \left(-\frac{1}{4}+\frac{\zeta_2}{2} \epsilon^2+\frac{5 \zeta_3}{2} \epsilon^3+\frac{11 \zeta_4}{4} \epsilon^4\right)
$}  \nn
\rlap{$\displaystyle\hspace{1cm}
+x^{-4\epsilon} \left(\frac{1}{8}+2 \zeta_2 \epsilon^2-\frac{\zeta_3}{2} \epsilon^3+56 \zeta_4 \epsilon^4\right)+\mathcal{O}(\epsilon^5), 
$}  \nn
g_{32}&\sim 
\rlap{$\displaystyle
\left(-\frac{3}{16}-\frac{3 \zeta_2}{4} \epsilon^2-\frac{3 \zeta_3}{2} \epsilon^3-12 \zeta_4 \epsilon^4\right)
+x^{-2\epsilon} \left(\frac{3}{8}-\frac{3 \zeta_2}{4} \epsilon^2-\frac{15 \zeta_3}{4} \epsilon^3-\frac{33 \zeta_4}{8} \epsilon^4\right)
$}  \nn
\rlap{$\displaystyle\hspace{1cm}
+x^{-4\epsilon} \left(\frac{1}{16}+\zeta_2 \epsilon^2-\frac{\zeta_3}{4} \epsilon^3+28 \zeta_4 \epsilon^4\right)+\mathcal{O}(\epsilon^5), 
$}  \nn
g_{33}&\sim 
\rlap{$\displaystyle
\left(-\frac{\zeta_2}{2} \epsilon^2-\frac{5 \zeta_3}{2} \epsilon^3-\frac{55 \zeta_4}{4} \epsilon^4\right)
+x^{-\epsilon} \left(\zeta_2 \epsilon^2+6 \zeta_3 \epsilon^3+\frac{81 \zeta_4}{4} \epsilon^4\right)
$}  \nn
\rlap{$\displaystyle\hspace{1cm}
+x^{-2\epsilon} \left(-\frac{\zeta_2}{2} \epsilon^2-\frac{7 \zeta_3}{2} \epsilon^3-\frac{25 \zeta_4}{2} \epsilon^4\right)+\mathcal{O}(\epsilon^5), 
$}  \nn
g_{34}&\sim \left(-\frac{\zeta_2}{2} \epsilon^2-\frac{5 \zeta_3}{2} \epsilon^3-\frac{55 \zeta_4}{4} \epsilon^4\right)
+x^{-2\epsilon} \left(\frac{\zeta_2}{2} \epsilon^2+\frac{7 \zeta_3}{2} \epsilon^3+\frac{25 \zeta_4}{2} \epsilon^4\right)+\mathcal{O}(\epsilon^5), \nn
g_{35}&\sim x^{-2\epsilon} \left(-1+3 \zeta_2 \epsilon^2+17 \zeta_3 \epsilon^3+36 \zeta_4 \epsilon^4\right)
+x^{-4\epsilon} \left(\frac{1}{4}+4 \zeta_2 \epsilon^2-\zeta_3 \epsilon^3+112 \zeta_4 \epsilon^4\right)+\mathcal{O}(\epsilon^5), \nn
g_{37}&\sim x^{-2\epsilon} \left(-\frac{3}{2}+3 \zeta_2 \epsilon^2+15 \zeta_3 \epsilon^3+\frac{33 \zeta_4}{2} \epsilon^4\right)
+x^{-3\epsilon} \left(\frac{1}{2}+\frac{5 \zeta_2}{2} \epsilon^2-4 \zeta_3 \epsilon^3+\frac{207 \zeta_4}{8} \epsilon^4\right)+\mathcal{O}(\epsilon^5), \nn
g_{38}&= \frac{1}{8}+\frac{\zeta_2}{2} \epsilon^2+\zeta_3 \epsilon^3+8 \zeta_4 \epsilon^4+\mathcal{O}(\epsilon^5), \nn
g_{39}&\sim x^{-2\epsilon} \left(-\frac{3}{4}+\frac{3 \zeta_2}{2} \epsilon^2+\frac{15 \zeta_3}{2} \epsilon^3+\frac{33 \zeta_4}{4} \epsilon^4\right)
+x^{-4\epsilon} \left(\frac{1}{8}+2 \zeta_2 \epsilon^2-\frac{\zeta_3}{2} \epsilon^3+56 \zeta_4 \epsilon^4\right)+\mathcal{O}(\epsilon^5), \nn
g_{43}&\sim -\frac{27 \zeta_4}{4} \epsilon^4+\mathcal{O}(\epsilon^5), \nn
g_{44}&\sim 
\rlap{$\displaystyle
\left(-\frac{5}{24}+\frac{\zeta_2}{3} \epsilon^2+\frac{7 \zeta_3}{6} \epsilon^3+\frac{\zeta_4}{2} \epsilon^4\right)
+x^{-2\epsilon} \left(\frac{3}{4}-\frac{3 \zeta_2}{2} \epsilon^2-\frac{15 \zeta_3}{2} \epsilon^3-\frac{33 \zeta_4}{4} \epsilon^4\right)
$}  \nn
\rlap{$\displaystyle
+x^{-3\epsilon} \left(-\frac{2}{3}-\frac{10 \zeta_2}{3} \epsilon^2+\frac{16 \zeta_3}{3} \epsilon^3-\frac{69 \zeta_4}{2} \epsilon^4\right)
+x^{-4\epsilon} \left(\frac{1}{8}+\frac{3 \zeta_2}{2} \epsilon^2-2 \zeta_3 \epsilon^3+\frac{63 \zeta_4}{2} \epsilon^4\right)+\mathcal{O}(\epsilon^5), 
$}  \nn
g_{45}&\sim 
\rlap{$\displaystyle
\left(-\frac{1}{24}-\frac{5 \zeta_2}{6} \epsilon^2-\frac{2 \zeta_3}{3} \epsilon^3-\frac{11 \zeta_4}{4} \epsilon^4\right)
+x^{-3\epsilon} \left(-\frac{1}{3}-\frac{5 \zeta_2}{3} \epsilon^2+\frac{8 \zeta_3}{3} \epsilon^3-\frac{69 \zeta_4}{4} \epsilon^4\right)
$}  \nn
\rlap{$\displaystyle\hspace{1cm}
+x^{-4\epsilon} \left(\frac{1}{8}+\frac{3 \zeta_2}{2} \epsilon^2-2 \zeta_3 \epsilon^3+\frac{63 \zeta_4}{2} \epsilon^4\right)+\mathcal{O}(\epsilon^5), 
$}  \nn
g_{46}&= -\frac{27 \zeta_4}{4} \epsilon^4+\mathcal{O}(\epsilon^5), \nn
g_{47}&= -\frac{\zeta_2}{4} \epsilon^2+\frac{\zeta_3}{4} \epsilon^3+9 \zeta_4 \epsilon^4+\mathcal{O}(\epsilon^5), \nn
g_{50}&\sim 
\rlap{$\displaystyle
\left(\frac{1}{12}+\frac{5 \zeta_2 \epsilon^2}{12}+\frac{31 \zeta_3 \epsilon^3}{12}+10 \zeta_4 \epsilon^4\right)
+x^{-2\epsilon} \left(-\frac{1}{4}-\zeta_3 \epsilon^3-\frac{39 \zeta_4}{4} \epsilon^4\right)
$}  \nn
\rlap{$\displaystyle
+x^{-3\epsilon} \left(\frac{1}{6}+\frac{5 \zeta_2}{6} \epsilon^2-\frac{4 \zeta_3}{3} \epsilon^3+\frac{69 \zeta_4}{8} \epsilon^4\right)
+x^{-4\epsilon} \left(\frac{\zeta_2}{4} \epsilon^2+\frac{3 \zeta_3}{4} \epsilon^3+\frac{49 \zeta_4}{4} \epsilon^4\right)+\mathcal{O}(\epsilon^5), 
$}  \nn
g_{51}&\sim \left(\frac{\zeta_2}{2} \epsilon^2+\frac{5 \zeta_3}{2} \epsilon^3+7 \zeta_4 \epsilon^4\right)
+x^{-2\epsilon} \left(-\frac{\zeta_2}{2} \epsilon^2-\frac{7 \zeta_3}{2} \epsilon^3-\frac{25 \zeta_4}{2} \epsilon^4\right)+\mathcal{O}(\epsilon^5), \nn
g_{52}&\sim 
\rlap{$\displaystyle
\left(-\frac{1}{12}-\frac{5 \zeta_2 \epsilon^2}{12}-\frac{31 \zeta_3 \epsilon^3}{12}-10 \zeta_4 \epsilon^4\right)
+x^{-2\epsilon} \left(\frac{1}{4}+\zeta_3 \epsilon^3+\frac{39 \zeta_4}{4} \epsilon^4\right)
$}  \nn
\rlap{$\displaystyle
+x^{-3\epsilon} \left(-\frac{1}{6}-\frac{5 \zeta_2}{6} \epsilon^2+\frac{4 \zeta_3}{3} \epsilon^3-\frac{69 \zeta_4}{8} \epsilon^4\right)
+x^{-4\epsilon} \left(\frac{1}{8}+\frac{9 \zeta_2}{4} \epsilon^2+\frac{\zeta_3}{4} \epsilon^3+\frac{273 \zeta_4}{4} \epsilon^4\right)+\mathcal{O}(\epsilon^5), 
$}  \nn
g_{54}&\sim 
\rlap{$\displaystyle
\left(-\frac{1}{6}-\frac{\zeta_2}{3} \epsilon^2+\frac{\zeta_3}{3} \epsilon^3-\frac{3 \zeta_4}{2} \epsilon^4\right)
+x^{-2\epsilon} \left(\frac{1}{2}-4 \zeta_3 \epsilon^3-6 \zeta_4 \epsilon^4\right)
$}  \nn
\rlap{$\displaystyle\hspace{1cm}
+x^{-3\epsilon} \left(-\frac{1}{3}-\frac{8 \zeta_2}{3} \epsilon^2+\frac{2 \zeta_3}{3} \epsilon^3-39 \zeta_4 \epsilon^4\right)+\mathcal{O}(\epsilon^5), 
$}  \nn
g_{55}&\sim x^{-2\epsilon} \left(-\frac{3}{4}+6 \zeta_3 \epsilon^3+9 \zeta_4 \epsilon^4\right)
+x^{-3\epsilon} \left(\frac{1}{3}+\frac{8 \zeta_2}{3} \epsilon^2-\frac{2 \zeta_3}{3} \epsilon^3+39 \zeta_4 \epsilon^4\right)+\mathcal{O}(\epsilon^5), \nn
g_{56}&\sim 
\rlap{$\displaystyle
x^{-2\epsilon} \left(-1-\frac{\zeta_2}{2} \epsilon^2-\frac{9 \zeta_3}{2} \epsilon^3+\frac{23 \zeta_4}{4} \epsilon^4\right)
+x^{-3\epsilon} \left(\frac{2}{3}+\frac{10 \zeta_2}{3} \epsilon^2+\frac{14 \zeta_3}{3} \epsilon^3+42 \zeta_4 \epsilon^4\right)
$}  \nn
\rlap{$\displaystyle\hspace{1cm}
+x^{-4\epsilon} \left(-\frac{1}{8}-\frac{3 \zeta_2}{2} \epsilon^2-31 \zeta_4 \epsilon^4\right)+\mathcal{O}(\epsilon^5), 
$}  \nn
g_{57}&\sim x^{-2\epsilon} \left(\frac{3}{4}-\frac{\zeta_2}{2} \epsilon^2-\frac{\zeta_3}{2} \epsilon^3-\frac{45 \zeta_4}{4} \epsilon^4\right)
+x^{-3\epsilon} \left(-\frac{1}{3}-\frac{5 \zeta_2}{3} \epsilon^2-\frac{7 \zeta_3}{3} \epsilon^3-21 \zeta_4 \epsilon^4\right)+\mathcal{O}(\epsilon^5), \nn
g_{58}&\sim 
\rlap{$\displaystyle
x^{-\epsilon} \left(\frac{2}{3}-\frac{5 \zeta_2}{3} \epsilon^2-\frac{22 \zeta_3}{3} \epsilon^3-\frac{87 \zeta_4}{4} \epsilon^4\right)
+x^{-2\epsilon} \left(-\frac{3}{2}+2 \zeta_2 \epsilon^2+8 \zeta_3 \epsilon^3+\frac{39 \zeta_4}{2} \epsilon^4\right)
$}  \nn
\rlap{$\displaystyle
+x^{-3\epsilon} \left(1+5 \zeta_2 \epsilon^2+2 \zeta_3 \epsilon^3+\frac{237 \zeta_4}{4} \epsilon^4\right)
+x^{-4\epsilon} \left(-\frac{1}{6}-\frac{7 \zeta_2}{3} \epsilon^2+\frac{\zeta_3}{3} \epsilon^3-58 \zeta_4 \epsilon^4\right)+\mathcal{O}(\epsilon^5), 
$}  \nn
g_{59}&\sim 
\rlap{$\displaystyle
x^{-2\epsilon} \left(-1+\frac{3 \zeta_2}{2} \epsilon^2+\frac{13 \zeta_3}{2} \epsilon^3-\frac{3 \zeta_4}{2} \epsilon^4\right)
+x^{-3\epsilon} \left(\frac{1}{3}+\frac{5 \zeta_2}{3} \epsilon^2-\frac{8 \zeta_3}{3} \epsilon^3+\frac{69 \zeta_4}{4} \epsilon^4\right)
$}  \nn
\rlap{$\displaystyle\hspace{1cm}
+x^{-4\epsilon} \left(\frac{\zeta_2}{4} \epsilon^2+\frac{3 \zeta_3}{4} \epsilon^3+\frac{49 \zeta_4}{4} \epsilon^4\right)+\mathcal{O}(\epsilon^5), 
$}  \nn
k_{19}&\sim -2 \zeta_3 \epsilon^3-\frac{7 \zeta_4}{2} \epsilon^4+\mathcal{O}(\epsilon^5), \numberthis \\
k_{20}&\sim \frac{\zeta_2}{2} \epsilon^2+\frac{5 \zeta_3}{2} \epsilon^3+\frac{15 \zeta_4}{2} \epsilon^4+\mathcal{O}(\epsilon^5), \nn
k_{21}&=-2 \zeta_3 \epsilon^3-\frac{7 \zeta_4}{2} \epsilon^4+\mathcal{O}(\epsilon^5), \nn
k_{22}&=\frac{\zeta_2}{2} \epsilon^2+\frac{5 \zeta_3}{2} \epsilon^3+\frac{15 \zeta_4}{2} \epsilon^4+\mathcal{O}(\epsilon^5), \nn
k_{26}&\sim \left(2 \zeta_3 \epsilon^3+\frac{39 \zeta_4}{4} \epsilon^4\right)+x^{-\epsilon} \left(-2 \zeta_3 \epsilon^3-12 \zeta_4 \epsilon^4\right)+\mathcal{O}(\epsilon^5), \nn
k_{27}&\sim \left(2 \zeta_2 \epsilon^2+4 \zeta_3 \epsilon^3+\frac{39 \zeta_4}{2} \epsilon^4\right)+x^{-\epsilon} \left(-\zeta_2 \epsilon^2-2 \zeta_3 \epsilon^3+\frac{15 \zeta_4}{4} \epsilon^4\right)+\mathcal{O}(\epsilon^5), \nn
k_{28}&\sim x^{-\epsilon} \left(\zeta_2 \epsilon^2+2 \zeta_3 \epsilon^3-\frac{15 \zeta_4}{4} \epsilon^4\right)+\mathcal{O}(\epsilon^5), \nn
k_{29}&=-\frac{\zeta_2}{2} \epsilon^2+\left(3 \zeta_2 \log (2)-\frac{7 \zeta_3}{4}\right)\epsilon^3 +\left(\frac{31 \zeta_4}{4}-6 \zeta_2 \log ^2(2)-12 \text{Li}_4\left(\frac{1}{2}\right)-\frac{\log ^4(2)}{2}\right)\epsilon^4 +\mathcal{O}(\epsilon^5), \nn
k_{30}&\sim -\zeta_3 \epsilon^3+\frac{3 \zeta_4}{4} \epsilon^4+\mathcal{O}(\epsilon^5), \nn
k_{31}&\sim -\zeta_2 \epsilon^2+\left(6 \zeta_2 \log (2)-\frac{9 \zeta_3}{2}\right)\epsilon^3 + \left(\frac{65 \zeta_4}{4}-12 \zeta_2 \log ^2(2)-24 \text{Li}_4\left(\frac{1}{2}\right)-\log ^4(2)\right)\epsilon^4+\mathcal{O}(\epsilon^5), \nn
k_{32}&\sim -\frac{\zeta_2}{2} \epsilon^2+\left(3 \zeta_2 \log (2)-\frac{9 \zeta_3}{4}\right)\epsilon^3 +\left(\frac{65 \zeta_4}{8}-6 \zeta_2 \log ^2(2)-12 \text{Li}_4\left(\frac{1}{2}\right)-\frac{\log ^4(2)}{2}\right)\epsilon^4 +\mathcal{O}(\epsilon^5), \nn
k_{33}&=-\zeta_3 \epsilon^3+\frac{3 \zeta_4}{4} \epsilon^4+\mathcal{O}(\epsilon^5), \nn
k_{34}&=\frac{\zeta_2}{4} \epsilon^2-\frac{3}{8}\left(4 \zeta_2 \log (2)-5 \zeta_3\right) \epsilon^3 +\left(-\frac{37 \zeta_4}{8}+3 \zeta_2 \log ^2(2)+6 \text{Li}_4\left(\frac{1}{2}\right)+\frac{\log ^4(2)}{4}\right)\epsilon^4 +\mathcal{O}(\epsilon^5), 
\end{align*}
and other MIs vanish in the limit we are considering. The results above can be regarded as the boundaries for the canonical differential equations at $s,t,m_W=0, m_t=1$. With the help of the package \texttt{DiffExp} \cite{Hidding:2020ytt}, we can obtain quite precise values for all MIs at a given kinematic point. Perfect agreement is found between the numerical results provided by \texttt{DiffExp} and our analytic evaluations.